\documentclass[aps,prd,twocolumn,flushleft,amsmath,amssymb,nofootinbib]{revtex4}
\usepackage{txfonts}
\usepackage{subfig}
\usepackage{graphicx}
\usepackage{dcolumn} 
\usepackage{bm}      
\usepackage[active]{srcltx} 
\usepackage{hyperref} 
\usepackage{amsmath}
\usepackage{amssymb}
\usepackage{stmaryrd}
\usepackage{amsfonts}
\usepackage{mathrsfs}

\newcommand{\be}{\begin{equation}}
\newcommand{\ee}{\end{equation}}
\newcommand{\ba}{\begin{eqnarray}}
\newcommand{\ea}{\end{eqnarray}}
\newcommand{\sgn}{\mathrm{sgn}} 
\newcommand{\Tr}{\mathrm{Tr}} 
\newcommand{\grav}{\mathrm{gr}} 
\newcommand{\sca}{\mathrm{sc}} 
\newcommand{\ints}{{\int_\Sigma}} 
\newcommand{\kin}{\mathrm{kin}} 
\newcommand{\hil}{\mathcal{H}} 
\newcommand{\Euc}{H^{Eucl}} 
\newcommand{\R}{\mathcal {R}} 
\newcommand{\abs}[1]{{\left|{#1}\right|}} 
\newcommand{\kt}{{\tilde{K}}} 
\newcommand{\bra}[1]{\langle{#1}\vert} 

\def\nn{\nonumber}

\begin{document}

\title{Loop quantum  $f(R)$ theories}

\author{Xiangdong Zhang\footnote{zhangxiangdong@mail.bnu.edu.cn} and Yongge Ma\footnote{mayg@bnu.edu.cn}}
\affiliation{Department of Physics, Beijing Normal University, Beijing 100875, China}

\begin{abstract}
As modified gravity theories, the 4-dimensional metric $f(\R)$
theories are cast into connection dynamical formalism with real
$su(2)$-connections as configuration variables. This formalism
enables us to extend the non-perturbative loop quantization scheme
of general relativity to any metric $f(\R)$ theories.  The quantum
kinematical framework of $f(\R)$ gravity is rigorously constructed,
where the quantum dynamics can be launched. Both Hamiltonian
constraint operator and master constraint operator for $f(\R)$
theories are well defined. Our results show that the
non-perturbative quantization procedure of loop quantum gravity are
valid not only for general relativity but also for a rather general
class of 4-dimensional metric theories of gravity.\\

\end{abstract}

\maketitle

\section{Introduction}\label{sec:introduction}

The theoretical search for a quantum theory of gravity has been
rather active. Especially, as a background independent approach to
quantize general relativity(GR), loop quantum gravity(LQG),  has
been widely investigated in recent twenty-five years. For reviews in
this field, we refer to\cite{Th07,Ro04,As04,Ma07}. It is remarkable
that, as a non-renormalizable theory, GR can be non-perturbatively
quantized by the loop quantization procedure\cite{Le10}. This
background-independent quantization relies on the key observation
that classical GR can be cast into the connection dynamical
formalism with structure group of $SU(2)$. Thus one is naturally led
to ask whether GR is a unique relativistic theory of gravity with
connection dynamical character. Recently modified gravity theories
have received increasinged attention in issues related to "dark
energy" and non-trivial tests on gravity beyond GR. A series of
independent observations, including type Ia supernova, weak lens,
cosmic microwave background anisotropy, baryon oscillation, etc,
implied that our universe is currently undergoing a period of
accelerated expansion\cite{01}. This result conflicts with the
prediction of GR and has carried the "dark energy" problem. Although
the acceleration could be explained by introducing a cosmological
constant $\Lambda$, the observed value of $\Lambda$ is unnaturally
much smaller than any estimation by tens of orders. Hence it is
reasonable to consider the possibility that GR is not a valid theory
of gravity on a cosmological scale. Since it was found that a small
modification  of the Einstein-Hilbert action by adding an inverse
power term of curvature scalar $\R$ would lead to current
acceleration of our universe, a large variety of models of $f(\R)$
modified gravity have been proposed\cite{So}. Moreover, some models
of $f(\R)$ gravity may account for the "dark matter" problem, which
was revealed by the observed rotation curve of galaxy clusters, We
refer to \cite{So,NO} for a recent review on $f(\R)$ theories of
gravity and It's application to cosmology. It is also worth noting
that certain effective equation of loop quantum cosmology can be
derived from some classical $f(\R)$ theory\cite{olmo09}.

Historically, Einstein's GR is the simplest relativistic theory of
gravity with correct Newtonian limit. It is worth pursuing all
alternatives, which provide a high chance to new physics. Recall
that the precession of Mercury's orbit was at first attributed to
some unobserved planet orbiting inside Mercury's orbit, but was
actually explained only after the passage from Newtonian gravity to
GR. Given the strong motivation  to $f(\R)$ gravity, it is desirable
to study such kind of theories at fundamental quantum level. For
metric $f(\R)$ theories, gravity is still geometry as in GR. The
differences between them are just reflected in dynamical equations.
Hence, a background-independent and non-perturbative quantization
for $f(\R)$ gravity is preferable. The framework of extending LQG to
$f(\R)$ theories appeared in \cite{Zh11}.The purpose of this paper
is to provide the detailed constructions.

We will show that the connection dynamical formulation of $f(\R)$
gravity can be derived by canonical transformations from it's
geometrical dynamics. The latter was realized by introducing a
non-minimally coupled scalar field to replace the original $f(\R)$
action and doing Hamiltonian analysis. While the equivalence by
canonical transformations at the classical level does not imply
equivalence after quantization, our choice of the canonical
formalism enables us to carry out the physical and mathematical
ideas of LQG. The canonical variables of our Hamiltonian formalism
of $f(\R)$ gravity consist of $su(2)$-connection $A_a^i$ and it's
conjugate momentum $E^a_i$ , as well as the scalar field $\phi$ and
it's momentum $\pi$. The Gaussian, diffeomorphism and Hamiltonian
constraints are also obtained, and they comprise a first-class
system. Loop quantization procedure is then naturally employed to
quantize $f(\R)$ gravity. The rigorous Kinematical Hilbert space
structure of loop quantum GR is extended to loop quantum $f(\R)$
gravity by adding a polymer-like quantum scalar field. The spatial
geometric operators of LQG, such as the area and volume operators
are still valid here. Hence the important physical result that both
the area and the volume are discrete at quantum kinematical level is also true
for $f(\R)$ gravity. As in LQG, the Gaussian and diffeomorphism
constraints can be solved at quantum level, and both the Hamiltonian
constraint and the master constraint can be promoted to well-defined
operators.

This paper is organized as follows. In section \ref{section1}, we
derive the connection dynamical formalism for $f(\R)$ theories. In
section \ref{section2}, the kinematical Hilbert space for $f(\R)$
gravity is constructed, where the Gaussian and diffeomorphism
constraints are implemented. The Hamiltonian constraint is promoted
to a well-defined operator in the kinematical Hilbert space in
section \ref{section3}. We also define a self-adjoint master
constraint operator in the diffeomorphism invariant Hilbert space in
section \ref{section4}. Finally, some concluding remarks are given
in section \ref{section5}. We use Greek alphabet for spacetime
indices. Latin alphabet a,b,c,...,for spatial indices, and
i,j,k,..., for internal indices.

\section{connection dynamical formalism for $f(R)$ theory}\label{section1}

A simple extension of GR is to consider the Lagrangian of gravity as
a function of scalar curvature $\R$. This kind of modified gravity
theories have become topical in cosmology and astro-physics. The
original action of $f(\R)$ theories read:

\ba S(g)=\frac12\int d^4x\sqrt{-g}f(\mathcal {R}) \label{action0}
\ea where $f$ is a general function of $\R$, and we set $8\pi G=1$.
By introducing an independent variable $s$ and a Lagrange multiplier
$\phi$, an equivalent action is proposed as\cite{Na,Na01}:
\ba S(g,\phi,s)&=&\frac12\int d^4x\sqrt{-g}(f(s)-\phi(s-\mathcal
{R})).\label{action} \ea The variation of (\ref{action}) with
respect to $s$ yields
\ba\phi=\frac{df(s)}{ds}\equiv f'(s).\ea
Assuming $f''(s)\neq 0$ so that $s$ could be
resolved from the above equation, action (\ref{action}) is reduced
to
\ba S(g,\phi) &=&\frac12\int d^4x\sqrt{-g}(\phi \mathcal
{R}-\xi(\phi))\equiv\int d^4x \mathcal {L}(x) \label{action1} \ea
where $\xi(\phi)\equiv\phi s-f(s)$. The variations of
(\ref{action1}) give the following equations of motion
\ba \phi
G_{\mu\nu}&=&-\frac12g_{\mu\nu}\xi(\phi)+\nabla_\mu\nabla_\nu\phi-g_{\mu\nu}\nabla_\sigma\nabla^\sigma\phi,\label{01}
\\
\mathcal {R}&=&\xi^\prime(\phi)\label{02}\ea
where
$\xi^\prime(\phi)\equiv\frac{d\xi(\phi)}{d\phi}$, and $\nabla_\mu$
is the connection compatible with $g_{\mu\nu}$. It is easy to see
that Eqs. (\ref{01}) and (\ref{02}) are equivalent to the equations
of motion derived from action (\ref{action0}). The virtue of action
(\ref{action1}) is that it admit a treatable Hamiltonian
analysis\cite{Na}. The Hamiltonian formalism can be derived by doing 3+1 decomposition and Legendre
transformation:
\ba p^{ab}&=&\frac{\partial\mathcal
{L}}{\partial\dot{h}_{ab}}\nn\\
&=&\frac{\sqrt{h}}{2}[\phi(K^{ab}-Kh^{ab})-\frac{h^{ab}}{N}(\dot{\phi}-N^c\partial_c\phi)], \label{04} \\
\pi&=&\frac{\partial\mathcal {L}}{\partial\dot{\phi}}=-\sqrt{h}K
\label{pi}\ea
where $h_{ab}$ and $K_{ab}$ are respectively the induced
3-metric and the extrinsic curvature of the spatial hypersurface
$\Sigma$ and $K\equiv K^a_a$. The combination of the trace of Eq.(\ref{04})
and Eq.(\ref{pi}) yields
\ba
\dot{\phi}-N^c\partial_c\phi=\frac{2N}{3\sqrt{h}}(\phi\pi-p).
\label{phidot}\ea
Note that the action (\ref{action1}) is also meaningful for a constant $\phi$. In this special case, one could resolve $\pi$ from $p$ by Eq.(\ref{phidot}) as $\pi=p/\phi$. This reduces one degree of freedom of the
theory. Then the $f(\R)$ theory will be reduced back to GR. In general case, the Hamiltonian of $f(\R)$ gravity can be derived as a liner
combination of constraints as
\ba H_{total}=\int_\Sigma d^3x(N^aV_a+NH).\label{htotal} \ea
where
$N$ and $N^a$ are the lapse function and shift vector respectively,
and the smeared diffeomorphism and Hamiltonian constraints read
\ba V(\overrightarrow{N})&\equiv&\int_\Sigma
d^3xN^aV_a\nn\\
&=&\int_\Sigma d^3xN^a(-2D^b(p_{ab})+\pi\partial_a\phi),\label{constraint1}\\
H(N)&\equiv&\int_\Sigma d^3xNH\nn\\
&=&\int_\Sigma d^3xN[\frac2{\sqrt{h}}(\frac{p_{ab}p^{ab}-\frac13p^2}{\phi}+\frac16\phi\pi^2-\frac13p\pi)\nn\\
&+&\frac12\sqrt{h}(\xi(\phi)-\phi
R+2D_aD^a\phi)],\label{constraint2} \ea
where $D_a$ is the
connection compatible with the 3-metric $h_{ab}$. Again, in the special case of
$\phi=costant$, it is easy
to see that the smeared diffeomorphism and Hamiltonian
constraints can go back to GR up to a constant rescale. By the
symplectic structure
\ba
\{h_{ab}(x),p^{cd}(y)\}&=&\delta^{(c}_a\delta^{d)}_b\delta^3(x,y),\nn\\
\{\phi(x),\pi(y)\}&=&\delta^3(x,y), \label{poission}\ea
lengthy but
straightforward calculations show that the constraints
(\ref{constraint1}) and (\ref{constraint2}) comprise a first class
system similar to GR as:
\ba
\{V(\overrightarrow{N}),V(\overrightarrow{N}^\prime)\}&=&V([\overrightarrow{N},\overrightarrow{N}^\prime]), \nn\\
\{V(\overrightarrow{N}),H(M)\}&=&H(\mathcal
{L}_{\overrightarrow{N}}M), \nn\\
\{H(N),H(M)\}&=&V(ND^aM-MD^aN). \ea
Since the above Hamiltonian
analysis is started with the action (\ref{action1}) where a
non-minimally coupled scalar field is introduced, we need to check
whether the Hamiltonian formalism is equivalent to the Lagrangian
formalism. It is not difficult to see
from the Hamiltonian (\ref{htotal}) that the evolution equation of
the scalar field reads:
\ba
\dot{\phi}=\{\phi,H_{total}\}=\frac{2N}{3\sqrt{h}}(\phi\pi-p)+N^a\partial_a\phi
\ea which is nothing but the Eq. of (\ref{phidot}). The evolution
equation of the 3-metric reads
\ba
\dot{h}_{ab}&=&\frac{N}{\sqrt{h}}(\frac{4(p_{ab}-\frac13ph_{ab})}{\phi}-\frac23\pi
h_{ab})+D_aN_b+D_bN_a\nn\\
&=&2NK_{ab}+D_aN_b+D_bN_a, \ea
which is nothing but the definition
of $K_{ab}$. The evolution equation of the momentum of $\phi$ reads
\ba \dot{\pi}
&=&\partial_a(N^a\pi)+\frac{2N}{\sqrt{h}}(\frac{p_{ab}p^{ab}-\frac13p^2}{\phi^2}-\frac16\pi^2)
-\frac{N\sqrt{h}}{2}\xi^\prime(\phi)\nn\\
&+&\frac{N\sqrt{h}}{2}R-\sqrt{h}D_aD^aN\nn\\
&=&\partial_a(N^a\pi)+\frac{N\sqrt{h}}{2}(K_{ab}K^{ab}-K^2)-\frac{N\sqrt{h}}{2}\xi^\prime(\phi)\nn\\
&+&\frac{N\sqrt{h}}{2}R-\partial_a(\sqrt{h}h^{ab}\partial_bN). \ea
Using the definition of $\pi=-\sqrt{h}K$ and
$n^0=\frac1N,n^i=-\frac{N^i}{N}$, we can get \ba
\xi^\prime(\phi)&=&K_{ab}K^{ab}-K^2+R+\frac{2}{N\sqrt{h}}\partial_\nu(\sqrt{-g}n^\nu
K)\nn\\
&-&\frac{2}{N\sqrt{h}}\partial_a(\sqrt{h}h^{ab}\partial_bN)=\mathcal
{R}. \ea This is nothing but Eq. (\ref{02}). On the other hand, the
00-component of Eq.(\ref{01}) reads \ba \phi G_{\mu\nu}n^\mu
n^\nu=\frac\phi2(R-K_{ab}K^{ab}+K^2).\label{eom00} \ea Using the
identity $g_{\mu\nu}n^\mu n^\nu=-1$, Eq.(\ref{eom00}) becomes \ba
&&\frac\phi2(R-K_{ab}K^{ab}+K^2)\nn\\
&=&\frac12\xi(\phi)+(g^{\mu\nu}+n^\mu
n^\nu)\nabla_\mu\nabla_\nu\phi\nn\\
&=&\frac12\xi(\phi)+D_aD^a\phi-K(\frac1N(\dot{\phi}-N^c\partial_c\phi))
\ea where the facts $h^{\mu\nu}n_\nu=0$ and
$n^\sigma\partial_\sigma\phi=\frac1N(\dot{\phi}-N^c\partial_c\phi)$
have been used in the above derivation. Note that the Hamiltonian
constraint in Eq.(\ref{constraint2}) can be expressed as 
\ba H
&=&\frac{\sqrt{h}\phi}{2}(K_{ab}K^{ab}-K^2-R)+\frac{\sqrt{h}}{2}(\xi(\phi)+2D_aD^a\phi)\nn\\
&-&\sqrt{h}K(\frac1N(\dot{\phi} -N^c\partial_c\phi)). \ea 
Hence the
00-component of (\ref{01}) is equivalent to Hamiltonian constraint.
Now we come to the 0a-component of (\ref{01}). Since \ba \phi
G_{\mu\nu}n^\mu h^\nu_a=\phi(D_aK^a_a-D_aK)  \quad and \quad
g_{\mu\nu}n^\mu h^\nu_a=0,\ea we have \ba
\phi(D_aK^a_b-D_bK)&=&n^\nu
h^\sigma_b\nabla_\sigma\nabla_\nu\phi\nn\\
&=&n^\nu h^\sigma_b\nabla_\sigma((h^\mu_\nu-n^\mu
n_\nu)\partial_\mu\phi). \ea The first term in the right hand side
of above equation reads \ba n^\nu
h^\sigma_a\nabla_\sigma(h^\mu_\nu\partial_\mu\phi)&=&n^\nu
h^\sigma_a\nabla_\sigma(g^\mu_\nu+n^\mu n_\nu)\partial_\mu\phi\nn\\
&=&-K^b_a\partial_b\phi, \ea and the second term reads \ba -n^\nu
h^\sigma_a\nabla_\sigma(n^\mu n_\nu\partial_\mu\phi)&=&h^\sigma_a
\nabla_\sigma(n^\mu\partial_\mu\phi)\nn\\
&=&D_a(\frac1N(\dot{\phi}-N^c\partial_c\phi)).\ea Hence their
combination gives \ba &&D_a(\phi K^a_b)-D_b(\phi K)+KD_b\phi
-D_b(\frac{1}{N}(\dot{\phi}-N^c\partial_c\phi))
\nn\\
&=&\frac{2}{\sqrt{h}}
D_a(\frac{\sqrt{h}}{2}[\phi(K^a_b-Kh^a_b)-\frac{h^a_b}{N}(\dot{\phi}-N^c\partial_c\phi)])\nn\\
&-&\frac{\pi}{\sqrt{h}}\partial_b\phi. \ea This is nothing but the
diffeomorphism constraint in Eq.(\ref{constraint1}). Now we turn to
the ab-components of (\ref{01}). We will show that they are
equivalent to the equation of motion of $p_{ab}$ which reads \ba
\dot{p}_{ab}&=&\frac{h_{ab}N}{\sqrt{h}}(\frac{p_{cd}p^{cd}-\frac13p^2}{\phi}+\frac16\phi\pi^2-\frac13p\pi)\nn\\
&+&\frac{2N}{\sqrt{h}}(\frac{p_{ac}p^c_b-\frac13pp_{ab}}{\phi}-\frac13p_{ab}\pi)\nn\\
&+&\frac N4\sqrt{h}h_{ab}\phi R-\frac N2\sqrt{h}\phi R_{ab}-\frac
N4\sqrt{h}h_{ab}\xi(\phi)\nn\\
&-&\frac
N2\sqrt{h}h_{ab}D_cD^c\phi-D_{(a}N\sqrt{h}D_{b)}\phi \nn\\
&+&\frac{\sqrt{h}}{2}(D_{(a}D_{b)}(N\phi)-h_{ab}D_cD^c(N\phi))\nn\\
&+&2p_{c(a}D^cN_{b)}+D_c(p_{ab}N^c).\label{pab} \ea Since the
initial value formalism of $f(\R)$ theories has been obtained in
\cite{So}, we will use Eq.(\ref{pab}) to derive the time derivative
of the extrinsic curvature: \ba
K_{ab}=\frac{2(p_{ab}-\frac13ph_{ab})}{\phi\sqrt{h}}-\frac{\pi
h_{ab}}{3\sqrt{h}}. \ea
A straightforward calculation yields
\ba \dot{K}_{ab}&=&2NK_{ac}K^c_b-NKK_{ab}+\mathcal
{L}_{\overrightarrow{N}}K_{ab}-NR_{ab}\nn\\
&+&D_aD_bN+\frac{N}{\phi}D_aD_b\phi\nn\\
&+&\frac{Nh_{ab}}{6}(\xi'(\phi)
+\frac{\xi(\phi)}{\phi})-\frac{n^\sigma\partial_\sigma\phi}{\phi}NK_{ab}.\label{kab}
\ea
It is easy to see that Eq.(\ref{kab}) is equivalent to Eq.(217)
in \cite{So}. Note that there is a sign difference between the
definition of our extrinsic curvature and that in \cite{So}, and our
potential $\xi(\phi)$ is twice of that in \cite{So}. To summarize,
we have shown that the Hamiltonian formalism of $f(\R)$ gravity is
equivalent to it's Lagrangian formalism.

Recall that the non-perturbative loop quantization of GR was based
on it's connection dynamic formalism. It is very interesting to
study whether the previous geometric dynamics of $f(\R)$ modified
gravity also has a connection dynamic correspondence. To this aim,
we first extend the phase space of geometrical dynamics to the triad formalism, and then introduce a canonical transformation on the extended
phase space of $f(\R)$ theories. Let
\ba
\tilde{K}^{ab}&\equiv&\phi
K^{ab}+\frac{h^{ab}}{2N}(\dot{\phi}-N^c\partial_c\phi)\nonumber\\
&=&\phi
K^{ab}+\frac{h^{ab}}{3\sqrt{h}}(\phi\pi-p),\label{ktilde}\ea
and
$E^a_i\equiv\sqrt{h}e^a_i$, where $e^a_i$ is the triad s.t.
$h_{ab}e^a_ie^b_j=\delta_{ij}$. Then we get
\ba
p^{ab}&=&\frac{\sqrt{h}}{2}(\tilde{K}^{ab}-\tilde{K}^c_ch^{ab}) \nn\\
&=&\frac{1}{2}(\tilde{K}^a_iE^{bi}-\frac1h\tilde{K}^i_cE^c_iE^a_jE^b_j), \nn\\
\pi&=&-\frac{\sqrt{h}}{\phi}(\tilde{K}^c_c-\frac{3}{2N}(\dot{\phi}-N^c\partial_c\phi)),
\ea
where $\tilde{K}^a_i\equiv\tilde{K}^{ab}e_b^i$.
Now we extend the phase space of geometry to the space consisting of pairs $(E^a_i, \tilde{K}_a^i)$. It is then easy to see that the
symplectic structure (\ref{poission}) can be derived from the following
Poisson brackets:
\ba
\{E^a_j(x),E^b_k(y)\}=\{\tilde{K}_a^j(x),\tilde{K}_b^k(y)\}=0,\nn\\
\{\tilde{K}^j_a(x),E_k^b(y)\}=\delta^b_a\delta^j_k\delta(x,y). \ea
Thus there is a symplectic reduction from the extended phase space to the original one, and the transformation from conjugate pairs $(h_{ab},p^{cd})$ to
$(E^a_i,\tilde{K}^j_b)$ is "canonical" in this sense. Note that since
$\tilde{K}^{ab}=\tilde{K}^{ba}$, we have an additional constraint:
\be G_{jk}\equiv\tilde{K}_{a[j}E^a_{k]}=0. \label{gaussian}\ee
So we
can further make a canonical transformation by defining:
\be A^i_a=\Gamma^i_a+\gamma\tilde{K}^i_a. \ee
where $\Gamma^i_a$ is
the spin connection determined by $E^a_i$, and $\gamma$ is a nonzero
real number. It is clear that our new variable $A^i_a$ coincides
with the Ashtekar-Barbero connection \cite{As86,Ba} when $\phi=1$.
The Poisson brackets among the new variables read:
\ba
\{A^j_a(x),E_k^b(y)\}&=&\gamma\delta^b_a\delta^j_k\delta(x,y),\nn\\
\{A_a^i(x),A_b^j(y)\}&=&0. \ea
Now, the phase space of $f(\R)$ gravity consists of
conjugate pairs $(A_a^i,E^b_j)$ and $(\phi,\pi)$. Combining
Eq.(\ref{gaussian}) with the compatibility condition:
\ba
\partial_aE^a_i+\epsilon_{ijk}\Gamma^j_aE^{ak}=0,
\ea
we obtain the standard Gaussian constraint
\ba \mathcal
{G}_i=\mathscr{D}_aE^a_i\equiv\partial_aE^a_i+\epsilon_{ijk}A^j_aE^{ak}
\label{GC}\ea
which justifies $A^i_a$ as an $su(2)$-connection. Note
that, had we let $\gamma=\pm i$, the (anti-)self-dual complex
connection formalism would be obtained. The original diffeomorphism
constraint can be expressed in terms of new variables up to Gaussian
constraint as
\ba V_a&=&-2D^b(p_{ab})+\pi\partial_a\phi\nn\\
&=&\frac1\gamma F^i_{ab}E^b_i+\pi\partial_a\phi, \ea
where
$F^i_{ab}\equiv2\partial_{[a}A^i_{b]}+\epsilon^i_{kl}A_a^kA_b^l$ is
the curvature of $A_a^i$. The original Hamiltonian constraint can be
written up to Gaussian constraint as
\ba
H&=&\frac{\phi}{2}[F^j_{ab}-(\gamma^2+\frac{1}{\phi^2})\varepsilon_{jmn}\tilde{K}^m_a\tilde{K}^n_b]
\frac{\varepsilon_{jkl}
E^a_kE^b_l}{\sqrt{h}}\nn\\
&+&\frac12(\frac2{3\phi}\frac{(\tilde{K}^i_aE^a_i)^2}{\sqrt{h}}+
\frac43\frac{(\tilde{K}^i_aE^a_i)\pi}{\sqrt{h}}+\frac23\frac{\pi^2\phi}{\sqrt{h}} \nn\\
&+&\sqrt{h}\xi(\phi))+\sqrt{h}D_aD^a\phi.\label{hamilton} \ea
It is
easy to check that the  smeared Gaussian constraint, $\mathcal
{G}(\Lambda):=\int_\Sigma d^3x\Lambda^i(x)G_i(x)$, generates $SU(2)$
gauge transformations on the phase space, while the smeared
constraint
\ba \mathcal {V}(\overrightarrow{N}):=\int_\Sigma
d^3xN^a(V_a-A_a^i\mathcal {G}_i) \ea
generates spatial
diffeomorphism transformations on the phase space. Together with the
smeared Hamiltonian constraint $H(N)=\int_\Sigma d^3xNH$, we can
show that the constraints algebra has the following form:
\ba \{\mathcal {G}(\Lambda),\mathcal
{G}(\Lambda^\prime)\}&=&\mathcal
{G}([\Lambda,\Lambda^\prime]),\label{eqsA} \\
\{\mathcal
{G}(\Lambda),\mathcal{V}(\overrightarrow{N})\}&=&-\mathcal{G}(\mathcal
{L}_{\overrightarrow{N}}\Lambda,),\\
\{\mathcal {G}(\Lambda),H(N)\}&=&0,\\
\{\mathcal {V}(\overrightarrow{N}),\mathcal
{V}(\overrightarrow{N}^\prime)\}&=&\mathcal
{V}([\overrightarrow{N},\overrightarrow{N}^\prime]), \\
\{\mathcal {V}(\overrightarrow{N}),H(M)\}&=&H(\mathcal
{L}_{\overrightarrow{N}}M),\label{eqsE}\\
\{H(N),H(M)\}&=&\mathcal {V}(ND^aM-MD^aN)\nn\\
&+&\mathcal
{G}\left((N\partial_aM-M\partial_aN)h^{ab}A_b\right)\nn\\
&-&\frac{[E^aD_aN,E^bD_bM]^i}{h}\mathcal {G}_i\nn\\
&-&\gamma^2\frac{[E^aD_a(\phi N),E^bD_b(\phi M)]^i}{h}\mathcal
{G}_i.\label{eqsb}\ea
One may understand Eqs.(\ref{eqsA}-\ref{eqsE}) by the geometrical interpretations of $\mathcal {G}(\Lambda)$ and
$\mathcal {V}(\overrightarrow{N})$. The detail calculation on the
Poisson bracket (\ref{eqsb}) between the two smeared Hamiltonian constraints will
be presented in the Appendix. Hence the constraints are of first
class. Moreover, the constraint algebra of GR can be recovered for
the special case when $\phi=1$. The total Hamiltonian is a linear
combination of the above constraints as \ba\hil_{tot}=\int_\Sigma
H(N)+N^aV_a+\mathcal {G}(\Lambda).\ea

To summarize, $f(\R)$ theories of gravity have been cast into the
$su(2)$-connection dynamical formalism. Though a scalar field is
non-minimally coupled, the resulted Hamiltonian structure is similar
to GR. Note that what we obtain is real $su(2)$-connection dynamics
of Lorentian $f(\R)$ gravity rather than the connection dynamics of
some conformal theories\cite{Fa10,Ci09}.

\section{Quantum kinematic of $f(\R)$ theory}\label{section2}

Recall that LQG is based on the connection dynamics of GR. We have
shown in last section that $f(\R)$ theories can also be reformulated
as connection dynamical theories by introducing a non-minimally
coupled scalar field. Hence the non-perturbative loop quantization
procedure can be straightforwardly generalized to $f(\R)$ theories.
Since the configuration space consists of geometry sector  and
scalar sector, we expect the kinematical Hilbert space of the system
to be an direct product of the Hilbert space of geometry and that of
scalar field. To construct quantum kinematics for geometry as in
LQG, we have to extend the space $\mathscr{A}$ of smooth connections
to space $\bar{\mathscr{A}}$ of distributional connections. A simple
element $\bar{A}\in\bar{\mathscr{A}}$ may be thought as a holonomy,
\ba h_e(A)=\mathcal {P}\exp\int_eA_a\ea of a connection along an
edge $e\subset\Sigma$. Through projective techniques,
$\bar{\mathscr{A}}$ is equipped with a natural measure $\mu_0$,
called the Ashtekar-Lewandowski measure\cite{As04,Ma07}. On the
other hand, one may smear the densitied triad $E^a_i$ on 2-surfaces
to obtain fluxes as \ba E(S,f):=\int_S\epsilon_{abc}E^a_if^i\ea
where $f^i$ is a $su(2)$-valued function on $S$. From the algebraic
viewpoint, the cylindrical functions of holonomies and the fluxes
consist of an $C*$-algebra. Then by Gel'fand-Naimark-Segal(GNS)
structure\cite{Th07}, one can obtain the cyclic representation for
the quantum holonomy-flux $*$-algebra, which coincides with the one
by projective techniques. In a certain sense, this is the unique
diffeomorphism and internal gauge invariant representation for the
quantum holonomy-flux algebra\cite{Le05}. The kinematical Hilbert
space of geometry then reads
$\hil^\grav_\kin=L^2(\bar{\mathscr{A}},d\mu_0)$. A typical vector
$\Psi_\alpha(\bar{A})\in\hil^\grav_\kin$ is a cylindrical function
over some finite graph $\alpha\subset\Sigma$. The so-called
spin-network basis \ba T_\alpha(\bar{A}) &=&\prod_{e\in
E(\alpha)}\sqrt{2j_e+1}\pi^{j_e}_{m_e,n_e}(\bar{A}(e)),\quad
(j_e\neq0)\nn\\\ea provides an orthonormal basis for
$\hil^\grav_\kin$\cite{As04,Ma07}, where
$\pi^{j_e}_{m_e,n_e}(\bar{A}(e))$ denotes the matrix elements in the
spin-$j$ representation of $SU(2)$. Note that the spatial geometric
operator of LQG, such as the area\cite{02} , the volume\cite{03} and
the length\cite{04,05} operators, are still valid in $\hil^\grav_\kin$, though their properties
in the physical Hilbert
space still need to be clarified \cite{Th09, Rovelli}.

Since the scalar field also reflects $f(\R)$ gravity, it is natural
to employ the polymer-like representation for it's quantization
\cite{06}. In this representation, one extends the space
$\mathscr{U}$ of smooth scalar fields to the quantum configuration
space $\bar{\mathscr{U}}$. A simple element
$\bar{U}\in\bar{\mathscr{U}}$ may be thought as a point holonomy,
\ba U_\lambda=\exp(i\lambda\phi(x)),\ea at point $x\in\Sigma$, where
$\lambda$ is a real number. By GNS structure\cite{Th07}, there is a
natural diffeomorphism invariant measure $d\mu$ on
$\bar{\mathscr{U}}$\cite{06}. Thus the kinematical Hilbert space of
scalar field reads $\hil^\sca_\kin=L^2(\bar{\mathscr{U}},d\mu)$. The
following scalar-network function of $\phi$:
\ba T_X(\phi)\equiv T_{X,\lambda}(\phi)=\prod_{x_j\in
X}U_\lambda(\phi(x_j)), \ea
where  $X=\{x_1,\dots, x_n\}$ is an
arbitrary given set of finite number of points in $\Sigma$,
constitute a complete set of orthonormal basis in
$\hil^{\sca}_\kin$. Since the point holonomy of a scalar is defined
on an 0-dimensional point, the momentum is smeared on an
3-dimensional region $R$ in $\Sigma$  as: \ba \pi(R):=\int_R
d^3x\pi(x).\ea Thus the total kinematical Hilbert space for $f(\R)$
gravity reads $\hil_\kin:=\hil^\grav_\kin\otimes \hil^\sca_\kin$
with an orthonormal basis $T_{\alpha,X}(A,\phi)\equiv
T_{\alpha}(A)\otimes T_{X}(\phi)$. Note that a basic feature of loop
quantization is that only holonomies will become configuration
operators, rather than the classical configuration variables
themselves. Let $\Psi(A,\phi)$ denote a quantum state in
$\hil_\kin$. The actions of basic operators read
\ba  \hat{h}_e(A)\Psi(A,\phi)&=&h_e(A)\Psi(A,\phi),
\nn\\
\hat{E}(S,f)\Psi(A,\phi)&=&i\hbar\{E(S,f),\Psi(A,\phi)\},\nn\\
\hat{U}_\lambda(\phi(v))\Psi(A,\phi)&=&\exp(i\lambda\phi(v))\Psi(A,\phi), \nn\\
\hat{\pi}(R)\Psi(A,\phi)&=&i\hbar\{\pi(R),\Psi(A,\phi)\}.  \ea
As in LQG, it is straight-forward to promote the Gaussian constraint
$\mathcal {G}(\Lambda)$ to a well-defined operator in $\hil_\kin$.
It's kernel is the internal gauge invariant Hilbert space $\mathcal
{H}_G$ with gauge invariant spin-scalar-network basis
$T_{s,c}=T_s(A)\otimes T_X(\phi)$, where \ba
T_{s=(\alpha,j,i)}(A)=\otimes_{v\in
V(\alpha)}i_v\cdot\otimes\pi^{j_e}(\bar{A}(e)),\quad(j_e\neq0). \ea
Here an intertwiner $i$ is assigned to each vertex of graph
$\alpha$. All the internal gauge invariant geometric operators, such
as the area, volume and length, can also be well defined in $\hil
_G$. Since the diffeomorphisms of $\Sigma$ act covariantly on the
cylindrical functions in $\mathcal {H}_G$, the so-called group
averaging technique can be employed to solve the diffeomorphism
constraint\cite{As04,Ma07}. To this aim, we first define a
projection map acting on cylindrical functions
$\psi_\beta\equiv\psi_{\alpha,X}(A,\phi)\in\hil_\kin$ as
\ba
\hat{P}_{Diff_\beta}\psi_\beta\coloneqq
\frac1{n_\beta}\sum_{\varphi\in
GS_\beta}\hat{U}_{\varphi}\psi_\beta, \ea
where $\hat{U}_{\varphi}$
denotes the unitary operator corresponding to a finite
diffeomorphism $\varphi : \Sigma\rightarrow \Sigma$. Here
$GS_\beta=Diff_\beta/TDiff_\beta$ is the group of graph symmetries,
where $Diff_\beta$ is the group of all diffeomorphisms preserving
the colored $\beta$, $TDiff_\beta$ is the group of diffeomorphisms
which trivially acts on $\beta$, and $n_\beta$ is the number of the
elements in $GS_\beta$. Secondly, we average with respect to all
remaining diffeomorphisms which change the graph $\beta$. For each
cylindrical function $\psi_\beta$, there is an element
$\eta(\psi_\beta)$ associated to it in the algebraic dual space
$Cyl^\star$ which acts on any cylindrical function $\phi_{\beta'}$
as
\ba \eta(\psi_\beta)[\phi_{\beta'}]\coloneqq\sum_{\varphi\in
Diff(\Sigma)/Diff_\beta}\bra{\hat{U}_\varphi\hat{P}_{Diff_\beta}\psi_\beta}\phi_{\beta'}\rangle_\kin,
\ea
where $Diff(\Sigma)$ is the diffeomorphism group of $\Sigma$. It
is easy to verify that $\eta(\psi_\beta)$ is invariant under the
group action of $Diff(\Sigma)$, since \ba
\eta(\psi_\beta)[\hat{U}_\varphi\phi_{\beta'}]=\eta(\psi_\beta)[\phi_{\beta'}].
\ea Thus we have defined a rigging map $\eta: Cyl\rightarrow
Cyl^\star_{Diff}$, which maps every cylindrical function to a
diffeomorphism invariant one. Moreover, a Hermitian inner product
can be defined on $Cyl^\star_{Diff}$ via the natural action of the
algebraic functional:
\ba
\bra{\eta(\psi_\beta)}\eta(\phi_{\beta'})\rangle_{Diff}\coloneqq\eta(\psi_\beta)[\phi_{\beta'}].
\ea
The diffeomorphism invariant Hilbert space $\hil_{Diff}$ is
defined by the completion of $Cyl^\star_{Diff}$ with respect to the
above inner product. Thus we can also obtain the desired
diffeomorphism and gauge invariant Hilbert space, $\mathcal
{H}_{Diff}$, for $f(\R)$ gravity.

\section{Quantum Hamiltonian of $f(\R)$ theory} \label{section3}

While the kinematical frameworks of LQG and polymer-like scalar
field have been straight-forwardly extended to $f(\R)$ theories, the
nontrivial task is to implement the Hamiltonian constraint
(\ref{hamilton}) at quantum level. In this section, we can show by
detail and technical analysis that, as in LQG, the Hamiltonian
constraint can be promoted to a well-defined operator in the
kinematical Hilbert space $\hil_\kin$. The resulted Hamiltonian
constraint operator is internal gauge invariant and diffeomorphism
covariant. Hence it is at least well defined in the gauge invariant
Hilbert space $\mathcal {H}_G$.

Comparing Eq.(\ref{hamilton}) with the Hamiltonian constraint of GR
in connection formalism, the new ingredient of $f(\R)$ gravity that
we have to deal with are $\phi(x),\phi^{-1}(x),\xi(\phi)$ and the
following four terms

\ba  H_3&=&\int_\Sigma d^3x\frac
N{3\phi}\frac{(\tilde{K}^i_aE^a_i)^2}{\sqrt{h}},\nn\\
H_4&=&\int_\Sigma d^3x \frac{2N}{3}
\frac{(\tilde{K}^i_aE^a_i)\pi}{\sqrt{h}},\nn\\
H_5&=&\int_\Sigma d^3x
\frac{N}{3}\frac{\pi^2\phi}{\sqrt{h}},\nn\\
H_7&=&\int_\Sigma d^3x N\sqrt{h}D_aD^a\phi. \label{newterm}\ea
Here
we have written the smeared version of Eq.(\ref{hamilton}) as
$H(N)=\sum^7_{i=1}H_i$. Note that the first two terms in $H(N)$ can
be written as \ba H_1&=&\frac{1}{2}\int_\Sigma d^3xN\phi
F^j_{ab}\frac{\varepsilon_{jkl}
E^a_kE^b_l}{\sqrt{h}} \nn\\
&=&\Euc(\phi N), \ea \ba H_2&=&-\frac{1}{2}\int_\Sigma
d^3xN(\gamma^2\phi+\frac{1}{\phi})\varepsilon_{jmn}\tilde{K}^m_a\tilde{K}^n_b\frac{\varepsilon_{jkl}
E^a_kE^b_l}{\sqrt{h}} \nn\\
&=&\frac{1}{1+\gamma^2}\mathcal {T}(N(\gamma^2\phi+\frac{1}{\phi})),
\ea where $\mathcal {T}(N)$ denotes the Lorentzian term in the
Hamiltonian constraint of GR. Hence, except that the smearing
functions are multiplied by some function of $\phi$, these terms
keep the same forms as those in GR.

By introducing certain small constant $\lambda_0$, an operator
corresponding to the scalar $\phi(x)$ at $x\in\Sigma$ can be defined
as
\ba
\hat{\phi}(x)=\frac{1}{2i\lambda_0}(U_{\lambda_0}(\phi(x))-U_{-\lambda_0}(\phi(x))).\label{phi}
\ea The ambiguity of $\lambda_0$ is the price that we have to pay in
order to represent field $\phi$ in the polymer-like representation.
To further define an operator corresponding to $\phi^{-1}(x)$, we
can use the classical identity \ba
\phi^{-1}(x)=\sgn[\phi](\frac1l\sgn[\phi]\{\abs{\phi}^{l}(x),\pi(R)\})^{\frac{1}{1-l}},\label{phi-1}
 \ea for any rational number $l\in(0,1)$, where $\sgn[\phi]$ denotes the sign function of $\phi$, $\abs{\phi}$ is the absolute value of $\phi$
and $x\in R$. For example, one may choose $l=\frac12$ for positive
$\phi(x)$ and replace the Poisson bracket by commutator to define
\ba
\hat{\phi}^{-1}(x)=(\frac{2}{i\hbar}[\sqrt{\hat{\phi}(x)},\hat{\pi}(R)])^2.
\ea Thus all the functions $\xi(\phi)$ which can be expanded as
powers of $\phi(x)$ have been quantized. For other non-trivial types
of $\xi(\phi)$, we may replace the argument $\phi$ by $\hat{\phi}$
in Eq.(\ref{phi}), provided that no divergence would arise after the
replacement. In the case where divergence does appear, there remain
the possibilities to employ tricks similar to Eq.(\ref{phi-1}) to
deal with it. Hence it is reasonable to believe that most physically
interesting functions $\xi(\phi)$ can be quantized. Then it is
straight-forward to quantize $H_6=\frac12\ints N\sqrt{h}\cdot
\xi(\phi)$ as an operator acting on an basis vector $T_{\alpha,X}$
as \ba \hat{H}_6\cdot T_{\alpha,X} &=&\frac12\sum_{v\in
V(\alpha)}N(v)\hat{\xi}(\phi(v))\hat{V_v}\cdot T_{\alpha,X}. \ea
Note that the action of the volume operator $\hat{V}$ on a
spin-network basis vector $T_\alpha(A)$ over a graph $\alpha$ can be
factorized as \ba \hat{V}\cdot T_\alpha=\sum_{v\in
V(\alpha)}\hat{V}_v\cdot T_\alpha.\ea Moreover, by the
regularization techniques developed for the Hamiltonian constraint
operators of LQG and polymer-like scalar field, all the terms
$H_3,H_4,H_5$ and $H_7$ can be regularized as operators acting on
cylindrical functions in $\hil_\kin$ in state-dependent ways. In the
regularization procedure. we will use the following classical
identities
\ba \tilde{K}^i_a=\frac{1}{\gamma}\{A_a^i,\kt\}, \ea
where
$\kt=\int_\Sigma d^3x\tilde{K}^i_aE^a_i$ can be write as Poisson
bracket:
\ba \kt=\gamma^{-\frac32}\{\Euc(1),V\}. \ea
Here the Euclidean scalar
constraint $\Euc(1)$ by definition was:
\ba \Euc(1)&=&\frac{1}{2}\int_\Sigma
d^3xF^j_{ab}\frac{\varepsilon_{jkl} E^a_kE^b_l}{\sqrt{h}}. \ea
Both
$\Euc$ and the volume $V$ under consideration have been quantized in
LQG. Also, one has
$E^a_i=\frac12\epsilon_{ijk}\epsilon^{abc}e_b^je_c^k$, where
$\epsilon^{abc}$ is the levi-civita tensor density, and the co-triad
satisfies \ba e^i_a=\frac2\gamma\{A^i_a(x),V\}.\ea To deal with the
four new terms (\ref{newterm}), we first regularize them separately
by point-splitting and obtain

\ba H_3 &=&\lim_{\epsilon\rightarrow 0}\int_\Sigma d^3y\int_\Sigma
d^3x\frac{N}{3\phi}\chi_\epsilon(x-y)\frac{\tilde{K}^i_a(x)E^a_i(x)}{\sqrt{V_{U^\epsilon_x}}}
\frac{\tilde{K}^j_b(y)E^b_j(y)}{\sqrt{V_{U^\epsilon_x}}}\nn\\
&=&\lim_{\epsilon\rightarrow
0}\frac{4}{3\gamma^3}\frac{N}{\phi}\chi_\epsilon(x-y)\{\Euc(1),\sqrt{(V_{U_x^\epsilon})}\}\nn\\
&&\{\Euc(1),\sqrt{(V_{U_y^\epsilon})}\}, \ea

\ba H_4 &=&\lim_{\epsilon\rightarrow
0}\frac{2^{15}}{3^4\gamma^6}\int_\Sigma
d^3y\pi(y)\chi_\epsilon(w-y)\nn\\
&\times&\int_\Sigma d^3xN\chi_\epsilon(x-y)\{A_a^i(x),\bar{K}\}
\nn\\
&\times&\epsilon^{abc}\Tr(\tau_i\{A_b(x),(V_{U_x^\epsilon})^{3/4}\}\{A_c(x),(V_{U_x^\epsilon})^{3/4}\}) \nn\\
&\times&\int_\Sigma
d^3w\epsilon^{def}\Tr(\{A_d(w),\sqrt{V_{U_w^\epsilon}}\}\{A_e(w),\sqrt{V_{U_w^\epsilon}}\}\nn\\
&\times&\{A_f(w),\sqrt{V_{U_w^\epsilon}}\}),  \ea

\ba H_5 &=&\lim_{\epsilon\rightarrow 0}\int_\Sigma d^3y\int_\Sigma
d^3x\frac{N\phi}{3}\chi_\epsilon(x-y)\pi(x)\pi(y) \nn\\
&\times&\int_\Sigma
d^3u\frac{det(e_a^i(u))}{(V^\epsilon_{U_u})^{3/2}}\chi_\epsilon(u-x)\nn\\
&\times&\int_\Sigma
d^3w\frac{det(e_a^i(w))}{(V^\epsilon_{U^\epsilon_w})^{3/2}}\chi_\epsilon(w-y)\nn\\
&=&\lim_{\epsilon\rightarrow 0}\frac{2^{14}}{3^3\gamma^6}\int_\Sigma
d^3y\int_\Sigma
d^3xN\phi\pi(x)\pi(y)\nn\\
&&\chi_\epsilon(x-y)\chi_\epsilon(u-x)\chi_\epsilon(w-y)
\nn\\
&\times&\int_\Sigma
d^3u\epsilon^{abc}\Tr(\{A_a(u),\sqrt{V_{U_u^\epsilon}}\}\nn\\
&&\{A_b(u),\sqrt{V_{U_u^\epsilon}}\}\{A_c(u),\sqrt{V_{U_u^\epsilon}}\}) \nn\\
&\times&\int_\Sigma
d^3w\epsilon^{def}\Tr(\{A_d(w),\sqrt{V_{U_w^\epsilon}}\}\nn\\
&&\{A_e(w),\sqrt{V_{U_w^\epsilon}}\}\{A_f(w),\sqrt{V_{U_w^\epsilon}}\}),
\ea

\ba H_7&=&-\int_\Sigma d^3xD_a(NE^a_i)\frac{1}{\sqrt{h}}E^b_iD_b\phi \nn\\
&=&\lim_{\epsilon\rightarrow 0}\int_\Sigma d^3y\int_\Sigma
d^3x(D_a\chi_\epsilon(x-y))
\nn\\
&\times&N(x)E^a_i(x)
\nn\\
&\times&(D_b\phi(y))\frac{E^b_i(y)}{\epsilon^{3}(h(y))^{1/2}}
\nn\\
&=&-\lim_{\epsilon\rightarrow 0}\frac{2^{5}}{\gamma^2}\int_\Sigma
d^3y\int_\Sigma d^3x(D_a\chi_\epsilon(x-y))N(x)E^a_i(x)
\nn\\
&\times&\epsilon^{bef}(D_b\phi(y))\Tr(\tau_i\{A_e(y),(V_{U_y^\epsilon})^{1/2}\}\{A_f(y),(V_{U_y^\epsilon})^{1/2}\}),
\nn\\\ea
where $\chi_\epsilon(x-y)$ is the characteristic function
of a box $U^\epsilon_x$ containing $x$ with scale $\epsilon$ and
satisfies the relation $\lim_{\epsilon\rightarrow
0}\chi_\epsilon(x-y)/\epsilon^3=\delta(x-y)$, and $V_{U^\epsilon_x}$
denote the volume of $U^\epsilon_x$. It is easy to see that the
regulator in $H_3$ can be removed by acting on a given basis vector
$T_{\alpha,X}\in\hil_\kin$ as \ba \hat{H}_3\cdot T_{\alpha,X} &=&
\sum_{v\in V(\alpha)}\frac{4N(v)}{3\gamma^3(i\hbar)^2}\hat{\phi}^{-1}(v)\nn\\
&\times& [\hat{\Euc}(1),\sqrt{\hat{V}_v}]
[\hat{\Euc}(1),\sqrt{\hat{V}_{v}}]\cdot T_{\alpha,X}, \nn\\
\ea For the other three terms, in order to reexpress connection by
holonomy and make the regularization diffeomorphism covariant, we
triangulate $\Sigma$ in adaptation to some graph $\alpha$ underling
a cylindrical function in $\hil_\kin$. At every vertex $v\in
V(\alpha)$, for each triple $(e_I,e_J,e_K)$ of edges of $\alpha$ we
have a tetrahedron $\Delta^\varepsilon_{\alpha,e_I,e_J,e_K}$ based
at $v$, which is spanned by segments $s_I,s_J,s_K$ of the triple.
Each segment $s_I$ is given by the part with the curve parameter
$t^I\in[0,\varepsilon]$ of the corresponding edge $e_I(t^I)$. For
each $\Delta^\varepsilon_{\alpha,e_I,e_J,e_K}$ one can construct
seven additional tetrahedron by backward analytic extension of the
segments. The regions of $\Sigma$ without a vertex of $\alpha$ can
be triangulated arbitrarily. Note that for one segment $s_I$, we
have
\begin{align}
\int_{s_I}\{A(u),\sqrt{V(u,\epsilon)}\}\approx\epsilon\dot{s}^a_I(0)\{A_a(v),\sqrt{V(u,\epsilon)\,}\}
\end{align}
up to $O(\epsilon^2)$. Hence for each
$\Delta_{\alpha,v,e_I,e_J,e_K}^\varepsilon$, we have \ba
&&\int_{\Delta_{\alpha,v,e_I,e_J,e_K}^\varepsilon}
    \epsilon^{abc}Tr\left(\left\{A_a(u),\sqrt{V_u^\epsilon}\;\right\}
    \left\{A_b(u),\sqrt{V_u^\epsilon}\;\right\}\left\{A_c(u),\sqrt{V_u^\epsilon}\;\right\}\right)\nonumber\\
&\approx&-\frac{1}{6}\epsilon(s_Is_Js_K)\epsilon^{IJK}\mathrm{Tr}\Big(h_{s_I(\Delta)}
\left\{h_{s_I(\Delta)}^{-1},\sqrt{V_{v(\Delta)}^\epsilon}\;\right\}
   \nonumber\\
&\times&h_{s_J(\Delta)}\left\{h_{s_J(\Delta)}^{-1},\sqrt{V_{v(\Delta)}^\epsilon}\;\right\}
h_{s_K(\Delta)}\left\{h_{s_K(\Delta)}^{-1},\sqrt{V_{v(\Delta)}^\epsilon}\;\right\}\Big),\nn\\
\ea where $\epsilon(s_I s_J
s_K):=\mathrm{sgn}(\det(\dot{s}_I\dot{s}_J\dot{s}_K)(v))$ takes the
values $+1, -1, 0$ if the tangents of the three segments
$s_I,s_J,s_K$ at $v$ (in that sequence) form a matrix of positive,
negative or vanishing determinant. Then the integration over
$\Sigma$ can be split as follows \cite{Th07}: \ba \label{triangle}
&&\int_\Sigma=\int_{\bar{U}_\alpha^\varepsilon}+\sum_{v\in
V(\alpha)}\int_{U^\varepsilon_{\alpha,
v}}\nn\\
&=&\int_{\bar{U}_\alpha^\varepsilon}+\sum_{v\in
V(\alpha)}\frac{1}{E(v)}\sum_{b(e_I)\cap b(e_J)\cap
b(e_K)=v}\left[\int_{U^\varepsilon_{\alpha,v,e_I,e_J
,e_K}}+\int_{\bar{U}^\varepsilon_{\alpha,v,e_I,e_J
,e_K}}\right]\nonumber\\
&\approx&\int_{\bar{U}_\alpha^\varepsilon}+\sum_{v\in
V(\alpha)}\frac{1}{E(v)}\sum_{b(e_I)\cap b(e_J)\cap
b(e_K)=v}\left[8\cdot\int_{\Delta^\varepsilon_{\alpha,v,e_I,e_J,
e_K}}+\int_{\bar{U}^\varepsilon_{\alpha,v,e_I,e_J,e_K}}\right].\nn\\
\ea
Here we have first decomposed $\Sigma$ into a region
$\bar{U}^\varepsilon_{\alpha}$ not containing the vertices of
$\alpha$ and the regions $U^\varepsilon_{\alpha, v}$ around the
vertices. Then choose a triple $(e_I, e_J, e_K)$ of edges outgoing
from $v$ and decompose $U^\varepsilon_{\alpha, v}$ into the region
$U^\varepsilon_{\alpha, v, e_I, e_J, e_K}$ covered by the
tetrahedron $\Delta^\varepsilon_{\alpha, v, e_I, e_J, e_K}$ spanned
by $e_I, e_J, e_K$ and its 7 mirror images and the rest
$\bar{U}^\varepsilon_{\alpha, v, e_I, e_J, e_K}$ not containing $v$.
Note that the integral over $U^\varepsilon_{\alpha, v, e_I, e_J,
e_K}$ classically converges to 8 times the integral over the
original single tetrahedron $\Delta^\varepsilon_{\alpha, v, e_I,
e_J, e_K}$ as we shrink the tetrahedron to zero. We average over all
such tripes $(e_I, e_J, e_K)$ and divide by the number of possible
choices of triples for a vertex $v$ with $n(v)$ edges,
$E(v)=\Big({n(v)\atop 3}\Big)$. Then by the above triangulation
$T(\varepsilon)$, the regulated 3 terms become respectively
\ba
H^\varepsilon_4 &=&-\lim_{\epsilon\rightarrow 0}\frac{2^{20}}{3^6\gamma^6}\ints d^3y\pi(y)\nn\\
&\times&\sum_{v\in\alpha(v)}\frac{N(v(\Delta))}{E(v)}\sum_{v(\Delta)=v}\chi_{\epsilon}(v(\Delta^{''})-y)
\chi_{\epsilon}(v(\Delta)-y))\nn\\
&\times&\Tr(\tau_i h_{s_L(\Delta)}\{h^{-1}_{s_L(\Delta)},\kt\})
\nn\\
&\times&\epsilon^{LMN}\epsilon(s_L s_M
s_N)\Tr(\tau_ih_{s_M(\Delta)}\{h^{-1}_{s_M(\Delta)},(V_{U_{v(\Delta)}^\epsilon})^{3/4}\}\nn\\
&\times&h_{s_N(\Delta)}\{h^{-1}_{s_N(\Delta)},(V_{U_{v(\Delta)}^\epsilon})^{3/4}\}) \nn\\
&\times&\sum_{v''\in\alpha(v)}\frac{1}{E(v'')}\sum_{v(\Delta)=v''}\epsilon(s_I
s_J
s_K)\epsilon^{IJK}\nn\\
&\times&\Tr(h_{s_I(\Delta'')}\{h^{-1}_{s_I(\Delta'')},\sqrt{V_{U_{v({\Delta''})}^\epsilon}}\}
h_{s_J(\Delta'')}\{h^{-1}_{s_J(\Delta'')},\sqrt{V_{U_{v({\Delta''})}^\epsilon}}\}\nn\\
&\times&h_{s_K(\Delta'')}\{h^{-1}_{s_K(\Delta'')},\sqrt{V_{U_{v(\Delta'')}^\epsilon}}\}),
\nn \ea

\ba H^\varepsilon_5 &=&\lim_{\epsilon\rightarrow
0}\frac{2^{17}}{3^5\gamma^6}\nn\\
&\times&
\ints d^3xN(x)\phi(x)\pi(x)\ints d^3y\pi(y) \nn\\
&\times&\chi_\epsilon(v(\Delta^{'''})-y)\chi_\epsilon(v(\Delta^{''})-x)\chi_\epsilon(x-y)\nn\\
\nn\\
&\times&\sum_{v''\in\alpha(v)}\frac{1}{E(v'')}\sum_{v(\Delta)=v''}
\epsilon(s_I s_J s_K)\epsilon^{IJK}\nn\\
&\times&\Tr(h_{s_I(\Delta^{''})}\{h^{-1}_{s_I(\Delta^{''})},\sqrt{V_{U_{\Delta^{''}}^\epsilon}}\}
h_{s_J(\Delta^{''})}\{h^{-1}_{s_J(\Delta^{''})},\sqrt{V_{U_{\Delta^{''}}^\epsilon}}\}\nn\\
&\times&
h_{s_K(\Delta^{''})}\{h^{-1}_{s_K(\Delta^{''})},\sqrt{V_{U_{\Delta^{''}}^\epsilon}}\}) \nn\\
&\times&\sum_{v'''\in\alpha(v)}\frac{1}{E(v''')}\sum_{v(\Delta)=v'''}
\epsilon(s_L s_M s_N)\epsilon^{LMN}\nn\\
&\times&\Tr(h_{s_L(\Delta^{'''})}\{h^{-1}_{s_L(\Delta^{'''})},\sqrt{V_{U_{\Delta^{'''}}^\epsilon}}\}
h_{s_M(\Delta^{'''})}\{h^{-1}_{s_M(\Delta^{'''})},\sqrt{V_{U_{\Delta^{'''}}^\epsilon}}\}\nn\\
&\times&
h_{s_N(\Delta^{'''})}\{h^{-1}_{s_N(\Delta^{'''})},\sqrt{V_{U_{\Delta^{'''}}^\epsilon}}\}),
\nn \ea

\ba H^\varepsilon_7 &=&-\lim_{\epsilon\rightarrow
0}\frac{2^{7}}{3\gamma^2i\lambda_0} \int_\Sigma
d^3x(D_a\chi_\epsilon(x-v'))N(x)E^a_i(x)
\nn\\
&\times&\sum_{v'\in\alpha(v)}\frac{1}{E(v')}\sum_{v(\Delta')=v'}\epsilon(s_I
s_J
s_K)\epsilon^{IJK}\nn\\
&\times&U^{-1}_{\lambda_0}(\phi(s_{s_I(\Delta')}))
[U_{\lambda_0}(\phi(t_{s_I(\Delta')}))-U_{\lambda_0}(\phi(s_{s_I(\Delta')}))]\nn\\
&\times&\Tr(\tau_ih_{s_J(\Delta')}\{h^{-1}_{s_J(\Delta')},(V_{U_{\Delta'}^\epsilon})^{1/2}\}
h_{s_K(\Delta')}\{h^{-1}_{s_K(\Delta')},(V_{U_{\Delta'}^\epsilon})^{1/2}\}), \nn\\
\ea
where $v(\Delta)$ and $s_I(\Delta)$ denotes a vertex and a
segment of a tetrahedron respectively, and $t_{s_I(\Delta)}$ (
$s_{s_I(\Delta)}$) denotes target point (starting point) of a
segment $s_I(\Delta)$. Note that the action of the operator $\pi(R)$
on a scalar-network basis vector $T_X(\phi)$ over a graph $X$ can be
factorized as \ba\hat{\pi}(R)\cdot T_X=\sum_{x_i\in X\cap
R}\hat{\pi}_{x_i}\cdot T_X.\ea Now every ingredient of
$H^\varepsilon_i$ has clearly quantum analogy, we can define the
corresponding operators acting on a basis vector $T_{\alpha,X}$ over
some graph $\alpha\cup X$ as
\ba \hat{H}^\varepsilon_4\cdot
T_{\alpha,X} &=&-\lim_{\epsilon\rightarrow 0}\frac{2^{20}}{3^6\gamma^6(i\hbar)^6}\nn\\
&\times&\sum_{v'\in X}\hat{\pi}(v')\chi_{\epsilon}(v''-v')\chi_{\epsilon}(v'-v)\nn\\
&\times&\sum_{v\in\alpha(v)}\frac{N(v)}{E(v)}\sum_{v(\Delta)=v}\Tr(\tau_i\hat{h}_{s_L(\Delta_{v})}
[\hat{h}^{-1}_{s_L(\Delta_{v})},\hat{\kt}])
\nn\\
&\times&\epsilon(s_L s_M s_N)\epsilon^{LMN}\Tr(\tau_i\hat{h}_{s_M(\Delta_{v})}[\hat{h}^{-1}_{s_M(\Delta_{v})},(\hat{V}_{U^\epsilon_v})^{3/4}]\nn\\
&\times&\hat{h}_{s_N(\Delta_{v})}[\hat{h}^{-1}_{s_N(\Delta_{v})},(\hat{V}_{U^\epsilon_v})^{3/4}]) \nn\\
&\times&\sum_{v''\in\alpha(v)}\frac{1}{E(v'')}\sum_{v(\Delta)=v''}\epsilon(s_I s_J s_K)\epsilon^{IJK}\nn\\
&\times&\Tr(\hat{h}_{s_I(\Delta_{v''})}[\hat{h}^{-1}_{s_I(\Delta_{v''})},(\hat{V}_{U^\epsilon_{v''}})^{1/2}]\nn\\
&\times&\hat{h}_{s_J(\Delta_{v''})}[\hat{h}^{-1}_{s_J(\Delta_{v''})},(\hat{V}_{U^\epsilon_{v''}})^{1/2}] \nn\\
&\times&\hat{h}_{s_K(\Delta_{v''})}[\hat{h}^{-1}_{s_K(\Delta_{v''})},(\hat{V}_{U^\epsilon_{v''}})^{1/2}])\cdot
T_{\alpha,X}, \ea

\ba \hat{H}^\varepsilon_5\cdot T_{\alpha,X}
&=&\lim_{\epsilon\rightarrow
0}\frac{2^{18}}{3^5\gamma^6(i\hbar)^6}\sum_{v\in X}\sum_{v'\in X}
\nn\\
&\times&\hat{\phi}(v)N(v)\hat{\pi}(v)\hat{\pi}(v')\chi_\epsilon(v'''-v')\chi_\epsilon(v''-v)\chi_\epsilon(v'-v) \nn\\
&\times&\sum_{v''\in\alpha(v)}\frac{1}{E(v'')}\sum_{v(\Delta)=v''}
\epsilon(s_I s_J s_K)\epsilon^{IJK}\nn\\
&\times&\Tr(\hat{h}_{s_I(\Delta_{v''})}[\hat{h}^{-1}_{s_I(\Delta_{v''})},(\hat{V}_{U^\epsilon_{v''}})^{1/2}]\nn\\
&\times&
\hat{h}_{s_J(\Delta_{v''})}[\hat{h}^{-1}_{s_J(\Delta_{v''})},(\hat{V}_{U^\epsilon_{v''}})^{1/2}] \nn\\
&\times&\hat{h}_{s_K(\Delta_{v''})}[\hat{h}^{-1}_{s_K(\Delta_{v''})},(\hat{V}_{U^\epsilon_{v''}})^{1/2}]) \nn\\
&\times&\sum_{v'''\in\alpha(v)}\frac{1}{E(v''')}\sum_{v(\Delta)=v'''}
\epsilon(s_L s_M s_N)\epsilon^{LMN}\nn\\
&\times&\Tr(\hat{h}_{s_L(\Delta_{v'''})}[\hat{h}^{-1}_{s_L(\Delta_{U^\epsilon_{v'''}})},(\hat{V}_{v'''})^{1/2}]\nn\\
&\times&
\hat{h}_{s_M(\Delta_{v'''})}[\hat{h}^{-1}_{s_M(\Delta_{v'''})},(\hat{V}_{U^\epsilon_{v'''}})^{1/2}] \nn\\
&\times&\hat{h}_{s_N(\Delta_{v'''})}[\hat{h}^{-1}_{s_N(\Delta_{v'''})},(\hat{V}_{U^\epsilon_{v'''}})^{1/2}])\cdot
T_{\alpha,X}, \ea

\ba \hat{H}^\varepsilon_7\cdot
T_{\alpha,X}&=&-\lim_{\epsilon\rightarrow
0}\frac{2^{7}}{3\gamma^2i\lambda_0}
\sum_{e\in E(\alpha)}X^i_e(t_{k-1})\lim_{n\to\infty} \sum_{k=1}^n\nn\\
&\times&[\chi_\epsilon(e(t_k)-v')-\chi_\epsilon(e(t_{k-1})-v')]N(e(t_{k-1}))
\nn\\
&\times&\sum_{v'\in\alpha(v)}\frac{1}{E(v')}\sum_{v(\Delta)=v'}
\epsilon(s_I s_J s_K)\epsilon^{IJK}\nn\\
&\times&\hat{U}^{-1}_{\lambda_0}(\phi(s_{s_I(\Delta_{v'})}))
[\hat{U}_{\lambda_0}(\phi(t_{s_I(\Delta_{v'})}))-\hat{U}_{\lambda_0}(\phi(s_{s_I(\Delta_{v'})}))]\nn\\
&\times&\Tr(\tau_i\hat{h}_{s_J(\Delta_{v'})}[\hat{h}^{-1}_{s_J(\Delta_{v'})},(\hat{V}_{U^\epsilon_{v'}})^{1/2}]\nn\\
&\times&
\hat{h}_{s_K(\Delta_{v'})}[\hat{h}^{-1}_{s_K(\Delta_{v'})},(\hat{V}_{U^\epsilon_{v'}})^{1/2}])\cdot
T_{\alpha,X},\ea
where $0=t_0<t_1<..<t_n=1$ is an arbitrary
partition of the interval $[0,1]$, $X^i_e(t):= [h_e(0,t)\tau_i
h_e(t,1)]_{AB}
\partial/\partial[h_e(0,1)]_{AB}$ (we denote $X^i_e:=X^i_e(0)$ in the
following), and $h_{s_I(\Delta_{v})}$ denotes the holonomy along the
segment $s_I$ starting from the vertex $v$ of tetrahedron $\Delta$.
On the other hand, for $\hat{H}^\varepsilon_7$, we perform the limit
$n\rightarrow\infty$, and $\epsilon\rightarrow 0$ in reversed order.
Keeping $n$ fixed, for small enough $\epsilon$, only the term with
$k =1$ in the sum survives provided that $s_I(0)=v'$, So for small
enough $\epsilon$, the above operator reduces to
\ba\hat{H}^\varepsilon_7\cdot
T_{\alpha,X}&=&\lim_{\epsilon\rightarrow
0}\frac{2^{7}}{3\gamma^2i\lambda_0(i\hbar)^2}
\nn\\
&\times&\sum_{e\in E(\alpha)}X^i_e(0)\chi_\epsilon(e(0)-v')N(e(0))\nn\\
&\times&\sum_{v'\in\alpha(v)}\frac{1}{E(v')}\sum_{v(\Delta)=v'}
\epsilon(s_I s_J s_K)\epsilon^{IJK}\nn\\
&\times&\hat{U}^{-1}_{\lambda_0}(\phi(s_{s_I(\Delta_{v'})}))
[\hat{U}_{\lambda_0}(\phi(t_{s_I(\Delta_{v'})}))-\hat{U}_{\lambda_0}(\phi(s_{s_I(\Delta_{v'})}))]\nn\\
&\times&\Tr(\tau_i\hat{h}_{s_J(\Delta_{v'})}[\hat{h}^{-1}_{s_J(\Delta_{v'})},(\hat{V}_{U^\epsilon_{v'}})^{1/2}]\nn\\
&\times&
\hat{h}_{s_K(\Delta_{v'})}[\hat{h}^{-1}_{s_K(\Delta_{v'})},(\hat{V}_{U^\epsilon_{v'}})^{1/2}])\cdot
T_{\alpha,X}. \ea Since the actions of $\hat{H}^\varepsilon_4$ and
$\hat{H}^\varepsilon_5$ are independent of $\varepsilon$, we can
take the limits and obtain \ba \hat{H}_4\cdot T_{\alpha,X}
&=&-\sum_{v\in V(\alpha)\cap
X}\frac{2^{20}N(v)}{3^6\gamma^6(i\hbar)^6E^2(v)}\hat{\pi}(v)
\nn\\
&\times&\sum_{v(\Delta)=v(\Delta')=v}\Tr(\tau_i\hat{h}_{s_L(\Delta)}[\hat{h}^{-1}_{s_L(\Delta)},\hat{\kt}])\nn\\
&\times&\epsilon(s_L s_M s_N)\epsilon^{LMN}\nn\\
&\times&\Tr(\tau_i\hat{h}_{s_M(\Delta)}[\hat{h}^{-1}_{s_M(\Delta)},(\hat{V}_{v})^{3/4}]
\hat{h}_{s_N(\Delta)}[\hat{h}^{-1}_{s_N(\Delta)},(\hat{V}_{v})^{3/4}]) \nn\\
&\times&\epsilon(s_I s_J s_K)\epsilon^{IJK}\nn\\
&\times&\Tr(\hat{h}_{s_I(\Delta')}[\hat{h}^{-1}_{s_I(\Delta')},(\hat{V}_{v})^{1/2}]
\hat{h}_{s_J(\Delta')}[\hat{h}^{-1}_{s_J(\Delta')},(\hat{V}_{v})^{1/2}] \nn\\
&\times&\hat{h}_{s_K(\Delta')}[\hat{h}^{-1}_{s_K(\Delta')},(\hat{V}_{v})^{1/2}])\cdot
T_{\alpha,X}, \ea

\ba \hat{H}_5\cdot T_{\alpha,X} &=&\sum_{v\in V(\alpha)\cap
X}\frac{2^{18}N(v)}{3^5\gamma^6(i\hbar)^6E^2(v)}
\hat{\pi}(v)\hat{\phi}(v)\hat{\pi}(v) \nn\\
&\times&\sum_{v(\Delta)=v(\Delta')=v}\epsilon(s_I s_J
s_K)\epsilon^{IJK}\nn\\
&\times&\Tr(\hat{h}_{s_I(\Delta)}
[\hat{h}^{-1}_{s_I(\Delta)},(\hat{V}_{v})^{1/2}]
\hat{h}_{s_J(\Delta)}[\hat{h}^{-1}_{s_J(\Delta)},(\hat{V}_{v})^{1/2}] \nn\\
&\times&\hat{h}_{s_K(\Delta)}[\hat{h}^{-1}_{s_K(\Delta)},(\hat{V}_{v})^{1/2}]) \nn\\
&\times&\epsilon(s_L s_M s_N)\epsilon^{LMN}\nn\\
&\times&\Tr(\hat{h}_{s_L(\Delta')}[\hat{h}^{-1}_{s_L(\Delta')},(\hat{V}_{v})^{1/2}]
\hat{h}_{s_M(\Delta')}[\hat{h}^{-1}_{s_M(\Delta')},(\hat{V}_{v})^{1/2}] \nn\\
&\times&\hat{h}_{s_N(\Delta')}[\hat{h}^{-1}_{s_N(\Delta')},(\hat{V}_{v})^{1/2}])\cdot
T_{\alpha,X}. \ea
However, it is easy to see that the action of
$\hat{H}^\varepsilon_7$ on $ T_{\alpha,X}$ is graph changing. It
adds a finite number of vertices at $t(s_I(v))=\varepsilon$ for
edges $e_I(t)$ starting from each high-valent vertex of $\alpha$. As
a result, the family of operators $\hat{H}^\varepsilon_7(N)$ fails
to be weakly convergent when $\varepsilon\rightarrow 0$. However,
due to the diffeomorphism covariant properties of the triangulation,
the limit operator can be well defined via the so-called uniform
Rovelli-Smolin topology induced by diffeomorphism-invariant states
$\Phi_{Diff}$ as:
\ba \Phi_{Diff}(\hat{H}_7\cdot
T_{\alpha,X})=\lim_{\varepsilon\rightarrow
0}(\Phi_{Diff}|\hat{H}^\varepsilon_{7}|T_{\alpha,X}\rangle. \ea It
is obviously that the limit is independent of $\varepsilon$. Hence
both the regulators $\epsilon$ and $\varepsilon$ can be removed. We
then have \ba \hat{H}_7\cdot T_{\alpha,X} &=&\sum_{v\in
V(\alpha)}\frac{2^{7}N(v)}{3\gamma^2i\lambda_0(i\hbar)^2E(v)}\nn\\
&\times&\sum_{e(0)=v}X^i_e\sum_{v(\Delta)=v}\epsilon(s_I s_J
s_K)\epsilon^{IJK}\nn\\
&\times&\hat{U}^{-1}_{\lambda_0}(\phi(s_{s_I(\Delta)}))
[\hat{U}_{\lambda_0}(\phi(t_{s_I(\Delta)}))-\hat{U}_{\lambda_0}(\phi(s_{s_I(\Delta)}))]\nn\\
&\times&\Tr(\tau_i\hat{h}_{s_J(\Delta)})[\hat{h}^{-1}_{s_J(\Delta)},(\hat{V}_{v})^{1/2}]\nn\\
&\times&
\hat{h}_{s_K(\Delta)}[\hat{h}^{-1}_{s_K(\Delta)},(\hat{V}_{v})^{1/2}])\cdot
T_{\alpha,X} . \ea Collecting all terms, the whole Hamiltonian
constraint can be quantized as a well-defined operator $\hat{H}(N)$
in $\hil_\kin$. The action of $\hat{H}(N)$ on $T_{\alpha,X}$ can be
factorized as \ba \hat{H}(N)\cdot T_{\alpha,X}=\sum_{v\in
V(\alpha)}\hat{H}(N)_v\cdot T_{\alpha,X}. \ea This operator is
internal gauge invariant and hence also well defined in $\hil_G$.
However, although $\hat{H}(N)$ can dually act on the diffeomorphism
invariant states, there is no guarantee for the resulted states to
be still diffeomorphism invariant.

\section{Master constraint operator}\label{section4}

Although the Hamiltonian constraint operator constructed in last
section is well defined in $\hil_{G}$, it is difficult to define it
directly on $\hil_{Diff}$. Moreover, the constraint algebra
(\ref{eqsA})-(\ref{eqsb}) do not form a Lie algebra. This might lead
to quantum anomaly after quantization. In order to avoid possible
quantum anomaly and find the physical Hilbert space, master
constraint programme was first introduced by Thiemann in
\cite{Th03}. We now apply this programme to quantum $f(\R)$ gravity.

By definition the master constraint of $f(\R)$ theories classically
reads
\ba \mathcal {M}:=\frac12\int_\Sigma
d^3x\frac{\abs{H(x)}^2}{\sqrt{h}}, \label{mcs}\ea where the
Hamiltonian constraint $H(x)$ was given by Eq.(\ref{hamilton}). It
is obvious that \ba \mathcal {M}=0 \Leftrightarrow H(N)=0
\quad\forall N(x). \ea However, now the constraints form a Lie
algebra since
\ba \{\mathcal {V}(\overrightarrow{N}),\mathcal
{V}(\overrightarrow{N}^\prime)\}&=&\mathcal
{V}([\overrightarrow{N},\overrightarrow{N}^\prime]), \nn\\
\{\mathcal {V}(\overrightarrow{N}),\mathcal {M}\}&=&0, \nn\\
\{\mathcal {M},\mathcal {M}\}&=&0, \ea where diffeomorphism
constraints nicely form an ideal. The master constraint can be
regulated via a point-splitting strategy \cite{08} as:
\ba \mathcal {M}^\epsilon=\frac12\int_\Sigma d^3y\int_\Sigma
d^3x\chi_\epsilon(x-y)\frac{{H(x)}}{\sqrt{V_{U^\epsilon_x}}}\frac{{H(y)}}{\sqrt{V_{U^\epsilon_y}}}.
\ea
Introducing a partition $\mathcal {P}$ of the 3-manifold
$\Sigma$ into cells $C$, we have an operator
$\hat{H}^\varepsilon_{C,\beta}$ acting on spin-scalar-network basis
$T_{s,c}$ in $\hil_G$ via a state-dependent triangulation,
\ba
\hat{H}^\varepsilon_{C,\alpha}\cdot T_{s,c}=\sum_{v\in
V(\alpha)}\chi_C(v)\hat{H}(N)^\varepsilon_v \cdot T_{s,c}\ea
where
$\alpha$ denotes the underlying graph of the spin-network state
$T_{s}$, and
\ba\hat{H}(N)^\varepsilon_v=\sum_{v(\Delta)=v}\hat{H}^{\varepsilon,\Delta}_{GR,v}+\sum^7_{i=3}
\hat{H}^\varepsilon_{i,v}, \ea
with
\ba
\hat{H}^{\varepsilon,\Delta}_{GR,v}&=&\frac{32\hat{\phi}(v)}{3i\hbar\gamma
E(v)}\epsilon(s_I s_J s_K)\epsilon^{IJK}
\Tr(\hat{h}^{-1}_{\alpha_{IJ}(\Delta)}\hat{h}_{s_K(\Delta)}[\hat{h}^{-1}_{s_K(\Delta)},
\sqrt{\hat{V}_{U^\epsilon_{v}}}])\nn\\
&-&\frac{64}{(i\hbar)^3\gamma^3E(v)}(\hat{\phi}^{-1}(v)+\gamma^2\hat{\phi}(v))\epsilon(s_I
s_J s_K)\epsilon^{IJK}
\nn\\
&\times&\Tr(\hat{h}_{s_I(\Delta)}[\hat{h}^{-1}_{s_I(\Delta)},\hat{\kt}]
\hat{h}_{s_J(\Delta)}[\hat{h}^{-1}_{s_J(\Delta)},\hat{\kt}]
\hat{h}_{s_K(\Delta)}[\hat{h}^{-1}_{s_K(\Delta)},\sqrt{\hat{V}_{U^\epsilon_{v}}}]),\nn\\
\ea
and
\ba \hat{H}_{3,v}^\varepsilon &=&
\frac{16N(v)}{3\gamma^3(i\hbar)^2}\hat{\phi}^{-1}(v)\nn\\
&\times&
[\hat{\Euc}(1),(\hat{V}_{U^\epsilon_{v}})^{1/4}][\hat{\Euc}(1),(\hat{V}_{U^\epsilon_{v}})^{1/4}], \nn\\
\ea

\ba \hat{H}^\varepsilon_{4,v}
&=&-\sum_{v(\Delta)=v(\Delta')=v(X)=v}\frac{2^{18}N(v)}{3^4\gamma^6(i\hbar)^6E^2(v)}\hat{\pi}(v)
\nn\\
&\times&\Tr(\tau_i\hat{h}_{s_L(\Delta)}[\hat{h}^{-1}_{s_L(\Delta)},\hat{\kt}])\nn\\
&\times&\epsilon(s_L s_M s_N)\epsilon^{LMN}\Tr(\tau_i\hat{h}_{s_M(\Delta)}[\hat{h}^{-1}_{s_M(\Delta)},(\hat{V}_{U^\epsilon_{v}})^{1/2}]\nn\\
&\times&\hat{h}_{s_N(\Delta)}[\hat{h}^{-1}_{s_N(\Delta)},(\hat{V}_{U^\epsilon_{v}})^{1/2}]) \nn\\
&\times&\epsilon(s_I
s_J s_K)\epsilon^{IJK}\Tr(\hat{h}_{s_I(\Delta')}[\hat{h}^{-1}_{s_I(\Delta')},(\hat{V}_{U^\epsilon_{v}})^{1/2}]\nn\\
&\times&
\hat{h}_{s_J(\Delta')}[\hat{h}^{-1}_{s_J(\Delta')},(\hat{V}_{U^\epsilon_{v}})^{1/2}] \nn\\
&\times&\hat{h}_{s_K(\Delta')}[\hat{h}^{-1}_{s_K(\Delta')},(\hat{V}_{U^\epsilon_{v}})^{1/2}]),\ea

\ba \hat{H}^\varepsilon_{5,v}
&=&\sum_{v(\Delta)=v(\Delta')=v(X)=v}\frac{2^{20}N(v)}{3^5\gamma^6(i\hbar)^6E^2(v)}
\hat{\pi}(v)\hat{\phi}(v)\hat{\pi}(v) \nn\\
&\times&\epsilon(s_I s_J s_K)\epsilon^{IJK}\Tr(\hat{h}_{s_I(\Delta)}
[\hat{h}^{-1}_{s_I(\Delta)},(\hat{V}_{U^\epsilon_{v}})^{1/4}]\nn\\
&\times&
\hat{h}_{s_J(\Delta)}[\hat{h}^{-1}_{s_J(\Delta)},(\hat{V}_{U^\epsilon_{v}})^{1/2}] \nn\\
&\times&\hat{h}_{s_K(\Delta)}[\hat{h}^{-1}_{s_K(\Delta)},(\hat{V}_{U^\epsilon_{v}})^{1/2}]) \nn\\
&\times&\epsilon(s_L s_M s_N)\epsilon^{LMN}\Tr(\hat{h}_{s_L(\Delta')}[\hat{h}^{-1}_{s_L(\Delta')},(\hat{V}_{U^\epsilon_{v}})^{1/4}]\nn\\
&\times&
\hat{h}_{s_M(\Delta')}[\hat{h}^{-1}_{s_M(\Delta')},(\hat{V}_{U^\epsilon_{v}})^{1/2}] \nn\\
&\times&\hat{h}_{s_N(\Delta')}[\hat{h}^{-1}_{s_N(\Delta')},(\hat{V}_{U^\epsilon_{v}})^{1/2}]), \nn\\
\ea

\ba \hat{H}^\varepsilon_{6,v}
&=&\frac12N(v)\hat{\xi}(\phi(v))\sqrt{\hat{V}_{U^\epsilon_{v}}}, \ea

\ba \hat{H}^\varepsilon_{7,v}
&=&\frac{2^{9}}{3\gamma^2i\lambda_0(i\hbar)^2E(v)}\nn\\
&\times&\sum_{e(0)=v}X^i_e\sum_{v(\Delta)=v}\nn\\
&\times&\epsilon(s_I s_J s_K)\epsilon^{IJK}\hat{U}^{-1}_{\lambda_0}(\phi(s_{s_I(\Delta)}))\nn\\
&\times&
[\hat{U}_{\lambda_0}(\phi(t_{s_I(\Delta)}))-\hat{U}_{\lambda_0}(\phi(s_{s_I(\Delta)}))]\nn\\
&\times&\Tr(\tau_i\hat{h}_{s_J(\Delta)}[\hat{h}^{-1}_{s_J(\Delta)},(\hat{V}_{U^\epsilon_{v}})^{1/4}]\nn\\
&\times&
\hat{h}_{s_K(\Delta)}[\hat{h}^{-1}_{s_K(\Delta)},(\hat{V}_{U^\epsilon_{v}})^{1/4}]).
\ea  Note that the family of operators
$\hat{H}^\varepsilon_{C,\alpha}$ are cylindrically consistent up to
diffeomorphism. So the inductive limit operator $\hat{H}_{C}$ is
densely defined in $\hil_G$ by the uniform Rovelli- Smolin topology.
Moreover, the adjoint operators of $\hat{H}^\varepsilon_{C,\alpha}$,
which are also cylindrically consistent up to diffeomorphism, read
\ba (\hat{H}^\varepsilon_{C,\alpha})^\dagger\cdot T_{s,c}=\sum_{v\in
V(\alpha)}\chi_C(v)(\hat{H}(N)^\varepsilon_v)^\dagger\cdot T_{s,c}
\ea The inductive limit operator, $(\hat{H}_{C})^\dagger$, of
$(\hat{H}^\varepsilon_{C,\alpha})^\dagger$ is adjoint to
$\hat{H}_C$. Then we could define master constraint operator
$\hat{\mathcal {M}}$ on diffeomorphism invariant states as
\ba( \hat{\mathcal {M}}\Phi_{Diff})\cdot T_{s,c}=\lim_{\mathcal
{P}\rightarrow\Sigma,\varepsilon,\varepsilon'\rightarrow
0}\Phi_{Diff}[\frac12\sum_{c\in\mathcal
{P}}\hat{H}^\varepsilon_C(\hat{H}^{\varepsilon'}_C)^\dagger \cdot T_{s,c}
]\nn\\ \ea
Note that our construction of $\hat{\mathcal {M}}$ is
qualitatively similar to that in \cite{Ma06}, although the
quantitative actions are different. Similar to those in \cite{Ma06}
we can prove the following properties of $\hat{\mathcal {M}}$.

(i) $\hat{\mathcal {M}}$ is diffeomorphism invariant, i.e., \ba
(\hat{U}^{'}_\varphi\hat{\mathcal
{M}}\Phi_{Diff})\cdot T_{s,c}&=&(\hat{\mathcal {M}}\Phi_{Diff})\cdot T_{s,c},\nn\ea
where $\hat{U}^{'}_\varphi$ is induced by the unitary operator in $\hil_G$ corresponding to a finite
diffeomorphism transformation $\varphi$.

(ii) For any given diffeomorphism invariant spin-scalar-network
state $T_{[s,c]}$, the norm $\|\hat{\mathcal {M}}T_{[s,c]}\|_{Diff}
$ is finite. So $\hat{\mathcal {M}}$ is densely defined in $\mathcal
{H}_{Diff}$.

(iii) $\hat{\mathcal {M}}$ is a positive and symmetric operator in
$\hil_{Diff}$ and hence admits a unique self-adjoint Friedrichs
extension.

In conclusion, there exists a positive and self-adjoint operator
$\hat{\mathcal {M}}$ on $\hil_{Diff}$ corresponding to the master
constraint (\ref{mcs}). It is then possible to obtain the physical
Hilbert space of $f(\R)$ gravity by the direct integral
decomposition of $\hil_{Diff}$ with respect to $\hat{\mathcal {M}}$.

\section{concluding remarks}\label{section5}

How to unify quantum mechanics with gravity theory is one of the core problems in modern physics. In recent
twenty-five years, LQG has made considerable progress in quantizing GR non-perturbatively and hence become a
fascinating candidate theory for quantum gravity. This background-independent quantization relies on the key
observation that classical GR can be cast into the connection-dynamical formalism with the structure group of
$SU(2)$. Due to this particular formalism, LQG was generally considered as a quantization scheme that applies
only to GR. This was taken by many researchers to be a limitation of the quantization scheme. The fact of being of general
applicability would therefore be significative for the general debate about quantum gravity. Especially, $f(\R)$ gravity theories have become topical in issues related to dark energy in cosmology
and non-trivial astronomic tests beyond GR. Hence, whether such modified gravity theories could be quantized non-perturbatively is itself an interesting question.

The main results of Ref.\cite{Zh11} and the current paper can be summarized as follows. (i) The connection dynamics of
$f(\R)$ gravity has been obtained by canonical transformations from
it's geometric dynamics. (ii) Based on the $su(2)$-connection dynamical formalism, the rigorous kinematical framework of LQG has been successfully extended to metric $f(\R)$ gravity theories by coupling with a polymer-like scalar field.
The important physical result that both the area and the volume are discrete at quantum kinematical level remains valid for $f(\R)$ gravity. (iii) While the Hamiltonian constraint is promoted to well-defined operator in the kinematical Hilbert space, the master constraint can be promoted to well-defined operator in the diffeomorphism invariant Hilbert space of loop quantum $f(\R)$ gravity. Thus, the non-perturbative loop quantization procedure is not only
valid for GR but also valid for a rather general class of
4-dimensional metric theories of gravity. Therefore, the achievements which have been obtained are in two fold. First, classical metric $f(\R)$
theories have been successfully quantized non-perturbatively. This guarantees the existence of $f(\R)$ theories
of gravity at fundamental quantum level. Secondly, the valid range of LQG has been considerably enlarged to
include a rather general class of metric theories.

It should be noticed that classically the scalar field $\phi$ characterize different
$f(\R)$ theories of gravity by $\phi=f'(\R)$. Thus for a given
$f(\R)$ theory, $\phi$ becomes a particular function of scalar
curvature $\R$ while the potential $\xi(\phi)$ is fixed. Hence our quantum $f(\R)$ gravity may be understood
as a class of quantum theories representing different choices of the
function $f(\R)$. Of course, there are still many aspects of the connection formalism and loop quantization of $f(\R)$ theories which deserve discovering. For examples, it is still desirable to find an action for
the connection dynamics of $f(\R)$ gravity. The semiclassical analysis of loop quantum $f(\R)$ theories is yet to be done. To further explore the physical contents of the loop quantum $f(\R)$ gravity, we would like to study its applications to cosmology and black holes in future works. Moreover, It is also desirable to quantize $f(\R)$ theories by covariant spin foam approach.

\begin{acknowledgements}

This work is supported by NSFC (No.10975017) and the Fundamental
Research Funds for the Central Universities.
\end{acknowledgements}

\section*{Appendix}
We use $(\kt^i_a,E^b_j)$ and $(\phi,\pi)$ as canonical variables to derive the
constraints algebra. By the first canonical transformation, The
Hamiltonian constraint (\ref{constraint2}) in section II can be
written as
\ba
H&=&\frac{\sqrt{h}}{2\phi}[(\tilde{K}_{ab}\tilde{K}^{ab}-\tilde{K}^2)-\phi^2R]-\frac{2}{3\sqrt{h}}(\pi
p-\frac{p^2}{\phi})\nn\\
&+&\frac13\frac{\pi^2\phi}{\sqrt{h}}+\frac12\sqrt{h}\xi(\phi)+\sqrt{h}D_aD^a\phi,\nn\\
&=&\frac{1}{2\sqrt{h}\phi}(\kt^i_aE^b_i\kt^j_bE^a_j-\frac13\kt^i_aE^a_i\kt^j_bE^b_j)
-\frac12\phi\sqrt{h}R+\frac12\sqrt{h}\xi(\phi)\nn\\
&+&\frac{1}{2\sqrt{h}}(\frac43\kt^i_aE^a_i\pi+\frac23\pi^2\phi)+\sqrt{h}D_aD^a\phi.\ea
To calculate the Poisson bracket between two smeared Hamiltonian
constraints, we notice that the non-vanishing
contributions come only form the terms which contain the derivative of
canonical variables. Those terms are $\ints d^3xN\sqrt{h}D_aD^a\phi$,
which contains both the derivative of $E^b_j$ and the derivative of $\phi$,
and $\ints-\frac12\phi N\sqrt{h}R$, which only contain the derivative of $E^b_j$. Hence we first use
$\{\phi(x),\pi(y)\}=\delta^3(x,y)$ to calculate
\ba &\{&\ints
N\sqrt{h}D_aD^a\phi,\ints\frac{M}{2\sqrt{h}}(\frac43\kt^i_aE^a_i\pi+\frac23\pi^2\phi)\}_{(\phi,\pi)}-M
\leftrightarrow N \nn\\
&=&\ints(MD_aD^aN-ND_aD^aM)(\frac23\pi\phi+\frac23\kt^i_bE^b_i)\nn\\
&=&\ints(ND^aM-MD^aN)D_a(\frac23\pi\phi+\frac23\kt^i_bE^b_i).\label{hstart}\ea
Note also that
\ba
N\sqrt{h}D_aD^a\phi=N\sqrt{h}h^{ab}(\partial_a\partial_b\phi-\Gamma^c_{ab}\partial_c\phi).
\ea
Since only  $\Gamma^c_{ab}$ contains the derivative of $E^a_i$
in above equation, we consider
\ba
&&N\sqrt{h}h^{ab}\Gamma^c_{ab}\partial_c\phi\nn\\
&=&\frac
N2\sqrt{h}h^{ab}(\partial_c\phi)(h^{cd}(-\partial_ah_{bd}-\partial_bh_{ad}+\partial_dh_{ab}))\nn\\
&=&\frac
N2\sqrt{h}(\partial_c\phi)(2\partial_ah^{ac}-h_{ab}\partial^ch^{ab})\nn\\
&=&\frac
N2\sqrt{h}(\partial_c\phi)(2\partial_a(\frac{E^a_iE^c_i}{h})-h_{ab}\partial^c(\frac{E^a_iE^b_i}{h})).
\ea
Therefore, we use $\{\tilde{K}^j_a(x),E_k^b(y)\}=\delta^b_a\delta^j_k\delta(x,y)$ to calculate
\ba &&\{\ints
N\sqrt{h}(\partial_c\phi)\partial_a(\frac{E^a_iE^c_i}{h}),\ints\frac{M}{2\sqrt{h}}
(\frac1\phi(\kt^l_dE^b_l\kt^j_bE^d_j\nn\\
&-&\frac13\kt^l_dE^d_l\kt^j_bE^b_j)+\frac43\kt^l_dE^d_l\pi)\}_{(\tilde{K},E)}-M
\leftrightarrow N \nn\\
&=&\ints\frac12M(\partial_aN)(D_c\phi)\frac{2E^c_i}{h}(\frac2\phi(E^b_i\kt^j_bE^a_j\nn\\
&-&\frac13E^a_i\kt^j_bE^b_j)
+\frac43E^a_i\pi))\nn\\
&+&\frac12M(\partial_aN)(D_c\phi)\frac{E^a_iE^c_i}{h}(-E^j_d)(\frac2\phi(E^b_j\kt^m_bE^d_m\nn\\
&-&\frac13E^d_j\kt^m_bE^b_m) +\frac43E^d_j\pi))-M
\leftrightarrow N \nn\\
\ea
and
\ba &&\{ \ints-\frac N2\sqrt{h}(\partial_c\phi)
h_{ae}\partial^c(\frac{E^a_iE^e_i}{h}),\ints\frac{M}{2\sqrt{h}}
(\frac1\phi(\kt^l_dE^b_l\kt^j_bE^d_j\nn\\
&-&\frac13\kt^l_dE^d_l\kt^j_bE^b_j)+\frac43\kt^l_dE^d_l\pi)\}_{(\tilde{K},E)}-M
\leftrightarrow N \nn\\
&=&\ints-\frac14M(\partial^cN)(D_c\phi) h_{ae}\frac{2E^e_i}{h}(\frac2\phi(E^b_i\kt^j_bE^a_j\nn\\
&-&\frac13E^a_i\kt^j_bE^b_j)+\frac43E^a_i\pi))\nn\\
&-&\frac14M(\partial_aN)(D_c\phi)\frac{E^a_iE^c_i}{h}(-3E^j_d)(\frac2\phi(E^b_j\kt^m_bE^d_m\nn\\
&-&\frac13E^d_j\kt^m_bE^b_m) +\frac43E^d_j\pi))-M
\leftrightarrow N. \nn\\
\ea
The combination of above two Poisson brackets equals to
\ba
&&\ints(ND^aM-MD^aN)(-\frac13\pi
D_a\phi\nn\\
&-&\frac2\phi(\kt^j_bE^c_jh_{ac}D^b\phi
-\frac13\kt^j_bE^b_jD_a\phi)). \ea The variation of the terms
containing a derivative  in $\ints-\frac12\phi N\sqrt{h}R$ reads \ba
&&\ints\frac12\sqrt{h}(-D^aD^b(\phi N)+h^{ab}D_cD^c(\phi N))\delta
h_{ab}\nn\\
&=&\ints\frac12\sqrt{h}(D_aD_b(\phi N)-h_{ab}D_cD^c(\phi
N))\delta h^{ab}\nn\\
&=&\ints\frac12\sqrt{h}(D_aD_b(\phi N)-h_{ab}D_cD^c(\phi N))\delta
(\frac{E^a_iE^b_i}{h}).\nn\\\ea Thus we have \ba
&\{&\ints-\frac12\phi
N\sqrt{h}R,\ints\frac{M}{2\sqrt{h}}(\frac1\phi(\kt^l_dE^e_l\kt^j_eE^d_j\nn\\
&-&\frac13\kt^j_dE^d_j\kt^m_eE^e_m) +\frac43\kt^l_dE^d_l\pi)\}
-M \leftrightarrow N \nn\\
&=&\ints-\frac14(MD_aD_b(\phi N)-h_{ab}MD_cD^c(\phi
N))\frac{2E^b_i}{h}\nn\\
&&(\frac2\phi(E^e_i\kt^j_eE^a_j-\frac13\kt^j_dE^d_jE^a_i)+\frac43E^a_i\pi)\nn\\
&-&\frac14(-2MD_cD^c(\phi
N))(-E^i_a)(\frac2\phi(E^e_i\kt^j_eE^a_j\nn\\
&-&\frac13\kt^j_dE^d_jE^a_i)+\frac43E^a_i\pi)-M
\leftrightarrow N \nn\\
&=&\ints-(MD_aD_b(\phi N)-h_{ab}M(D_cD^c\phi
N))h^{be}\frac1\phi\kt^j_eE^a_j\nn\\
&-&M(D_cD^c\phi N)(\frac{2}{3\phi}\kt^j_dE^d_j+\frac23\pi) -M
\leftrightarrow N \nn\\
&=&\ints-M(D_aD^b\phi N)\frac1\phi\kt^j_bE^a_j +M(D_cD^c\phi
N)\nn\\
&&(\frac{1}{3\phi}\kt^j_dE^d_j-\frac23\pi) -M
\leftrightarrow N \nn\\
&=&\ints(ND_aD^b(\phi M)-MD_aD^b(\phi
N))\frac1\phi\kt^j_bE^a_j\nn\\
&+& (ND_cD^c(\phi M)-MD_cD^c(\phi
N))(\frac23\pi-\frac{1}{3\phi}\kt^j_dE^d_j)\nn\\
&=&\ints(ND_cD^cM-MD_cD^cN)(\frac23\pi\phi-\frac{1}{3}\kt^j_aE^a_j)\nn\\
&+&(ND_cM-MD_cN)(D^c\phi)(\frac43\pi-\frac{2}{3\phi}\kt^j_aE^a_j)\nn\\
&+&(ND_aD^bM-MD_aD^bN)\kt^j_bE^a_j\nn\\
&+&(ND_aM-MD_aN)\frac{2D^b\phi}{\phi}\kt^j_bE^a_j.\label{hend} \ea
Taking account of Eqs.(\ref{hstart})-(\ref{hend}), we obtain
\ba&&\{H(N),H(M)\}=\nn\\
&&\ints(ND_cD^cM-MD_cD^cN)(-\kt^j_aE^a_j)\nn\\
&+&(ND^aM-MD^aN)(\pi
D_a\phi)\nn\\
&+&(ND_aD^bM-MD_aD^bN)\kt^j_bE^a_j \nn\\
&=&\ints(ND^aM-MD^aN)(D_a(\kt^j_cE^c_j)-D_b(\kt^j_aE^b_j)+\pi
D_a\phi)\nn\\
&-&((D_aM)D^bN-(D^bM)D_aN)\kt^j_bE^a_j\nn\\
&=&\ints(ND^aM-MD^aN)V_a-\frac{[E^aD_aN,E^bD_bM]^i}{h}\mathcal
{G}_i\ea where we used the following identity \ba
&&-((D_aM)D^bN-(D^bM)D_aN)\kt^j_bE^a_j\nn\\
&=&-((D_aM)D_cN-(D_cM)D_aN)h^{b
c}E^a_j\kt^j_b\nn\\
&=&-2(D_{[a}M)(D_{c]}N)\frac{E^b_iE^c_i}{h}E^a_j\kt^j_b \nn\\
&=&-2(D_{a}M)(D_{c}N)\frac{E^{[a}_jE^{c]}_i}{h}\kt^j_bE^{ib}\nn\\
&=&-\epsilon^{ijk}(D_{a}M)(D_{c}N)\frac{E^{a}_{j}E^{c}_{i}}{h}\kt^m_bE^{nb}\varepsilon_{kmn}\nn\\
&=&-\frac{[E^aD_aN,E^bD_bM]^k}{h}\mathcal {G}_k.\ea
Using above
result and shift conjugate pair $(\kt^i_a,E^b_j)$ to
$(A^i_a,E^b_j)$, we can easily get the Poisson bracket (\ref{eqsb}) between the smeared
Hamiltonian constraints.


\newcommand\AL[3]{~Astron. Lett.{\bf ~#1}, #2~ (#3)}
\newcommand\AP[3]{~Astropart. Phys.{\bf ~#1}, #2~ (#3)}
\newcommand\AJ[3]{~Astron. J.{\bf ~#1}, #2~(#3)}
\newcommand\APJ[3]{~Astrophys. J.{\bf ~#1}, #2~ (#3)}
\newcommand\APJL[3]{~Astrophys. J. Lett. {\bf ~#1}, L#2~(#3)}
\newcommand\APJS[3]{~Astrophys. J. Suppl. Ser.{\bf ~#1}, #2~(#3)}
\newcommand\JCAP[3]{~JCAP. {\bf ~#1}, #2~ (#3)}
\newcommand\LRR[3]{~Living Rev. Relativity. {\bf ~#1}, #2~ (#3)}
\newcommand\MNRAS[3]{~Mon. Not. R. Astron. Soc.{\bf ~#1}, #2~(#3)}
\newcommand\MNRASL[3]{~Mon. Not. R. Astron. Soc.{\bf ~#1}, L#2~(#3)}
\newcommand\NPB[3]{~Nucl. Phys. B{\bf ~#1}, #2~(#3)}
\newcommand\PLB[3]{~Phys. Lett. B{\bf ~#1}, #2~(#3)}
\newcommand\PRL[3]{~Phys. Rev. Lett.{\bf ~#1}, #2~(#3)}
\newcommand\PR[3]{~Phys. Rep.{\bf ~#1}, #2~(#3)}
\newcommand\PRD[3]{~Phys. Rev. D{\bf ~#1}, #2~(#3)}
\newcommand\SJNP[3]{~Sov. J. Nucl. Phys.{\bf ~#1}, #2~(#3)}
\newcommand\ZPC[3]{~Z. Phys. C{\bf ~#1}, #2~(#3)}
\newcommand\CQG[3]{~Class. Quant. Grav. {\bf ~#1}, #2~(#3)}

\end{document}